\documentclass[aps,prb,twocolumn,showpacs,superscriptaddress,floatfix,10pt]{revtex4-1}
\usepackage{framed}
\usepackage{graphicx}
\usepackage{amsmath}
\usepackage{epsfig}
\usepackage{helvet}
\usepackage{amssymb}
\usepackage{amsthm}
\usepackage{color}
\usepackage{enumerate}
\usepackage[scriptsize]{subfigure}
\usepackage{times}
\usepackage{dsfont}
\usepackage{bbm}
\usepackage{mathrsfs}

\def\bra#1{\mathinner{\left\langle{#1}\right|}}
\def\ket#1{\mathinner{\left|{#1}\right\rangle}}
\def\braket#1{\mathinner{\left\langle{#1}\right\rangle}}

\def\bravert{\egroup\,\vrule\,\bgroup}

\renewcommand{\H}{\mathcal{H}}

\newcommand{\si}{\sigma}
\newcommand{\ep}{\epsilon}
\newcommand{\dg}{\dagger}

\newcommand{\F}{\mathcal{F}}
\newcommand{\mm}{\mathbf{m}}
\newcommand{\pp}{\mathbf{p}}
\newcommand{\rr}{\mathbf{r}}
\newcommand{\kk}{\mathbf{k}}
\newcommand{\qq}{\mathbf{q}}

\newcommand{\sss}{\mathbf{s}}

\newcommand{\id}{\mathds{1}}

\newcommand{\V}{\mathcal{V}}

\begin{document}

\title{The BCS wave function, matrix product states, and the Ising conformal field theory}

\author{Sebasti{\' a}n Montes}
\email{s.montes@csic.es}
\affiliation{Instituto de F\'{i}sica Te\'orica (IFT), UAM-CSIC, Madrid, Spain}

\author{Javier Rodr\'{\i}guez-Laguna}
\affiliation{Dept. of Fundamental Physics, Universidad Nacional de
  Educaci\'on a Distancia (UNED), Madrid, Spain}

\author{Germ\'an Sierra}
\affiliation{Instituto de F\'{i}sica Te\'orica (IFT), UAM-CSIC, Madrid, Spain}

\date{\today}

\begin{abstract} 
We present a characterization of the many-body lattice wave functions obtained from the conformal blocks (CBs) of the Ising conformal field theory (CFT). The formalism is interpreted as a matrix product state using continuous ancillary degrees of freedom. We provide analytic and numerical evidence that the resulting states can be written as BCS states. We give a complete proof that the translationally invariant 1D configurations have a BCS form and we find suitable parent Hamiltonians. In particular, we prove that the ground state of the finite-size critical Ising transverse field (ITF) Hamiltonian can be obtained with this construction. Finally, we study 2D configurations using an operator product expansion (OPE) approximation. We associate these states to the weak pairing phase of the $p+ip$ superconductor via the scaling of the pairing function and the entanglement spectrum.
\end{abstract}



\maketitle

\section{Introduction}

Even though superconductivity was discovered experimentally in 1911 by Kamerlingh Onnes \cite{Onnes}, a sufficiently predictive microscopic theory was not available until the work of Bardeen, Cooper and Schriefer, published in 1957 \cite{BCSoriginal} (known today as BCS theory). One of the fundamental features of this construction is the realization of the ground state of the system as a grand canonical state of fermionic pairs. Following these steps, several other many-body systems have benefited from these insights and extended the result to other non-trivial Gaussian states.

Another seminal landmark of many-body physics is Onsager's solution of the two-dimensional Ising model, published more than a decade earlier \cite{Onsager}. Despite its cumbersome original formulation, current understanding of this model is closely related to BCS theory. More concretely, the ground-state of the associated one-dimensional quantum spin chain can also be constructed from a condensate of fermionic pairs \cite{Henkel, Sachdev, Mussardo}. This is a remarkable result if we consider that both models have very different descriptions and applications.

BCS theory has remained an important starting point for the analysis of more exotic phenomena. For instance, in the past few decades two-dimensional superconductors have become testbeds for novel topological features, some of them closely related to the fractional quantum Hall (FQH) effect. Read and Green \cite{ReadGreen, Miguel} established a connection between the weak pairing regime of the $p+ip$ superconductor and the topological phase defined by the Moore-Read Pfaffian state \cite{MooreRead91}. The robustness of these phases to local perturbations have turned them into strong candidate schemes for quantum computing \cite{NayakAnyons}.

One of the shared tools for the study of these strongly correlated quantum systems is conformal field theory (CFT) \cite{BPZ, Tsvelik, QuantumGroupsCFT, diFrancesco, Gogolin, Mussardo}. Due to its symmetry constraints, these theories have powerful algebraic structures that may allow for exact solutions. This has been exploited in the construction of trial wave functions for many-body systems, both in the lattice and the continuum. This is done by computing correlators in the CFT and using them as variational wave functions. The most famous applications have been in FQH physics \cite{MooreRead91}, however it has also been used to study 1D spin systems using infinite matrix product states (iMPS) \cite{iMPS, iMPS2, KL, FQH-MPS}. In this latter case, the entanglement structure has some features that cannot be easily obtained from finite matrices, such as logarithmic scaling of the entanglement entropy \cite{iMPS}.

It has been argued that conformal blocks (CBs) of rational CFTs can be used to construct wave functions for lattice spin systems \cite{WFfromCB}. Even if there is no straightforward spin-like structure arising from the representation of internal symmetries (for instance, in the case of minimal CFTs \cite{BPZ}), the physical degrees of freedom can still be encoded in the different fusion channels of non-Abelian operators. This was illustrated using the Ising CFT, where the relevant CBs were obtained from chiral correlators of several spin operators $\sigma$, grouped in pairs to describe two-level systems.

In this paper, we provide further characterization of the many-body lattice wave functions obtained from the Ising CFT. In particular, we show both analytic and numerical evidence that states describing $N$ spins obtained from the CBs of $2N$ $\sigma$ fields (dubbed $\ket{\psi_{ee}}$) can be understood as BCS wave functions.

This article is organized as follows: Sect. II presents a general short review of BCS states. Sect. III and Sect. IV introduce the notion of vertex operators. We use this formalism to write $\ket{\psi_{ee}}$ and other related states as matrix product states with continuous ancillary degrees of freedom. In Sect. V, we develop a first-order operator product expansion (OPE) that allows us to write $\ket{\psi_{ee}}$ as an explicit (albeit approximate) BCS state. Sect. VI reviews the exact formulas for the Ising CBs. In Sect. VII, Sect. VIII and Sect. IX, we study in detail translationally invariant 1D states. We prove that in this case $\ket{\psi_{ee}}$ can be written as a BCS state in an exact manner and find suitable parent Hamiltonians. In particular, we prove that the ground-state of the finite critical Ising transverse field (ITF) spin chain corresponds to $\ket{\psi_{ee}}$ for a homogeneous configuration. In Sect. X, we study 1D excitations by means of wave functions obtained from CBs with different asymptotic boundary conditions. Sect. XI presents some general remarks about the problem of writing $\ket{\psi_{ee}}$ as a BCS state for an arbitrary coordinate configuration. Finally, in Sect. XII, we provide a brief study of the 2D states obtained from the OPE regime. We use both the entanglement spectrum \cite{LiHaldane} and the scaling of the entanglement entropy to relate these states to the weak pairing phase of the $p+ip$ superconductor.

%


\section{BCS wave functions: a short review}

Given a collection of (spinless) fermionic modes $\{c_n\}_{n=1}^N$ on a lattice, we can define the BCS many-body wave function
\begin{equation}
\ket{\psi_\text{BCS}} = \prod_{n<m}\left(u_{nm}+v_{nm}c_n^\dg c_m^\dg\right)\ket{0}_c,
\end{equation}
where $\ket{0}_c$ is the state annihilated by all the operators $c_n$, and $u_{nm}$, $v_{nm}$ are complex numbers that satisfy the normalization condition $|u_{nm}|^2+|v_{nm}|^2 = 1$. Furthermore, we impose $u_{nm}=u_{mn}$ and $v_{nm}=-v_{mn}$. This state can be written as
\begin{equation}
\ket{\psi_\text{BCS}} = C_N \exp\left(\sum_{n<m}g_{nm}c_n^\dg c_m^\dg \right)\ket{0}_c,
\end{equation}
where $g_{nm}=v_{nm}/u_{nm}$ is the pairing function (or more generally pairing matrix) and $C_N=\prod_{n<m} u_{nm}$ is a normalization constant. Note that $g_{nm}$ is a (generally complex) antisymmetric tensor $g_{nm}=-g_{mn}$.

We can interpret this wave function as a grand canonical state of pairs created by the operator $P=\sum_{n<m}g_{nm}c_n^\dg c_m^\dg$. From the fermionic anticommutation relations, it can be shown that the wave function amplitude for $2M$ fermions occupying sites $r(1)<\cdots<r(2M)$ is given by
\begin{equation}
\Psi(r(1),\cdots,r(2M))= C_N \text{Pf}(\textbf{M}),
\end{equation}
where $\textbf{M}$ is the $2M\times 2M$ antisymmetric matrix
\begin{equation}
(\textbf{M})_{ij} = g_{r(i), r(j)},
\end{equation}
and we make use of the Pfaffian
\begin{equation}
\text{Pf}(\textbf{Q}) = \frac{1}{2^{M}M!}\sum_{\sigma\in S_{2M}} \text{sgn}(\sigma) \prod_{j=1}^M (\textbf{Q})_{\sigma(2j-1),\sigma(2j) }.
\label{Pfaffian}
\end{equation}

BCS wave functions are Gaussian states that arise naturally from mean-field solutions of Hamiltonians describing superconductivity \cite{Sachdev, ReadGreen}. In that context, both $u_{ij}$ and $v_{ij}$ can be written in terms of single-particle energies $\epsilon_k$ and the pairing interaction potential $V_{k,k'}$. Being spinless fermions, we say that these states correspond to $p$-wave superconductivity, due to the fact that the wave functions for the spatial degrees of freedom are antisymmetric.

The aim of this paper is to provide an alternative route to the BCS state using CFT. More precisely, we shall consider states obtained from the chiral conformal blocks of the critical Ising model and relate them to known BCS states.


\section{Vertex operators in the chiral Ising CFT}

The (chiral) Ising CFT is a minimal RCFT that consists of three primary fields, $\id$, $\chi$ (Majorana), and $\sigma$ (spin) with conformal weights $0$, $1/2$, and $1/16$, respectively. They have the (non-trivial) fusion rules \cite{diFrancesco}
\begin{equation}
\sigma\times\sigma = \id + \chi, \qquad \chi\times\chi = \id, \qquad \sigma\times\chi = \sigma.
\end{equation}
A conformal block (CB) in a RCFT is a chiral correlator that encodes an allowed fusion channel for a given set of primary fields. If we start with $N$ primaries $\{\phi_{j_n}\}$, a CB can be written as \cite{QuantumGroupsCFT, MooreSeiberg, MooreSeibergNotes}
\begin{equation}
\F_\pp(z_1,\cdots,z_N) = \braket{\prod_{n=1}^N \phi_{j_n}(z_n)}_\pp.
\end{equation}
where  $\pp$ labels the internal channels. The number of conformal blocks of this type depends on the possible allowed fusion channels of the $\phi_{j_n}$ fields.

The exact formulas for the CBs obtained from the Ising primary fields have been calculated in Ref.\cite{NayakWilczek,IsingCB}. We will be interested mainly in CBs containing only $2N$ spin field operators
\begin{equation}
\F^{(2N)}_\pp (z_1,\cdots,z_{2N}) = \braket{\si(z_1)\cdots\si(z_{2N})}_\pp.
\label{sigmaCBs}
\end{equation}
Given the fusion rules, a pair of $\sigma$ fields can be seen as a single degree of freedom \cite{WFfromCB}. This allows us to write the fusion channels in terms of local binary variables. In order to see this explicitly, we group the fields in reference pairs $[\sigma(z_{2n-1}),\sigma(z_{2n})]$. When they are fused pairwise, the different channels can be labeled using the vector $\mm=(m_1,\cdots,m_N)$, with $m_i=0$ ($1$) representing an identity operator $\id$ (a fermion $\chi$). In this representation, there is an enforced parity coming from the preservation of fermion parity, so that $\sum_i m_i \equiv 0 (\text{mod } 2)$.

\begin{figure}[ht]
  \centering

\includegraphics[width=1.0\linewidth]{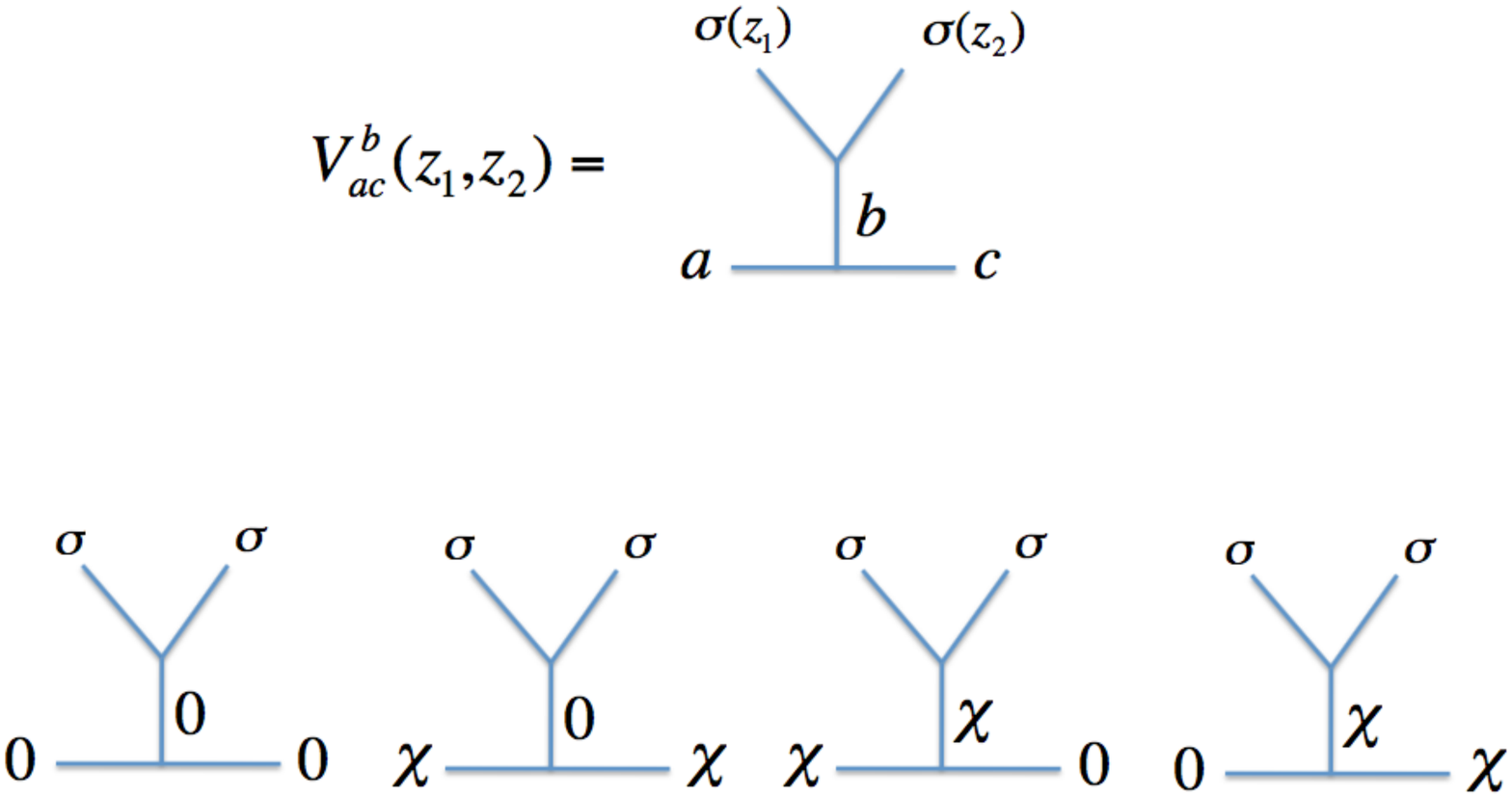}

    \caption{Graphical representation of the bilocal vertex operator $V^b_{ac}(z_1,z_2)$.}
    \label{bilocalvertex}
\end{figure}

The local pair-wise fusion produces bilocal chiral vertex operators
\begin{equation}
V_{ac}^b(z_{2n-1},z_{2n}) : \V_c \to \V_a, \qquad a,b,c = \id, \chi
\end{equation}
where $b=m_n$ corresponds to the fusion channel of reference pair $[\sigma(z_{2n-1}),\sigma(z_{2n})]$, $\V_a$ are the Verma modules associated to the corresponding primary fields, and we require the conservation of fermionic parity at each vertex [Fig.\eqref{bilocalvertex}]. We can use these operators to express explicitly the inner structure of each CB.

%


\section{Many-body lattice states from Ising conformal blocks}

Let us now consider the $2^N$-dimensional Hilbert space $\H$ obtained from $N$ spinless fermionic modes $\{c_n\}_{n=1}^N$ and define the $2\times 2$ operator matrix
\begin{equation}
A^{(n)}(z_{2n-1}, z_{2n}) = 
 \begin{pmatrix}
  V_{\id \id}^\id & c_n^\dg V_{\id\chi}^\chi \\
  c_n^\dg V_{\chi \id}^\chi & V_{\chi \chi}^\id 
 \end{pmatrix}.
 \label{Amatrix}
\end{equation}
This yields the map
\begin{equation}
A^{(n)}(z_{2n-1},z_{2n}) : \left(
    \begin{array}{c}
      \V_\id\otimes \H_e \\
      \V_\chi \otimes \H_o
    \end{array}
  \right)  \to  \left(
    \begin{array}{c}
      \V_\id\otimes \H_e \\
      \V_\chi \otimes \H_o
    \end{array}
  \right),  
\end{equation}
where $\H_e$ ($\H_o$) is the Fock space with even (odd) number of fermions
\begin{align}
\H_e &= \left\{\ket{0}_c, c_{i_1}^\dg c_{i_2}^\dg\ket{0}_c,\cdots\right\}, \\
\H_o &= \left\{ c_{i_1}^\dg\ket{0}_c, c_{i_1}^\dg c_{i_2}^\dg  c_{i_3}^\dg\ket{0}_c,\cdots\right\}, \nonumber
\end{align}
so that $\H=\H_e\oplus\H_o$. The product of $N$ matrices of type $A$ gives the $2\times 2$ operator matrix
\begin{align}
\Phi^{(N)} &= A^{(1)}(z_1,z_2)\cdots A^{(N)}(z_{2N-1},z_{2N})\nonumber\\
&=  \begin{pmatrix}
  \Phi_{ee}^{(N)} & \Phi_{eo}^{(N)} \\
   \Phi_{oe}^{(N)} & \Phi_{oo}^{(N)}
 \end{pmatrix}.
 \label{PsiMatrix}
\end{align}
Using this notation, we have that the operator $\Phi_{ee}^{(N)}$ acting on $\V_\id\otimes\H_e$ defines the (unnormalized) state
\begin{equation}
\ket{\psi_{ee}} = \braket{ 0\left| \Phi_{ee}^{(N)} \right| 0} \ket{0}_c \in\H_e,
\label{psiee}
\end{equation}
where $\bra{0}\cdots \ket{0}$ corresponds to the expectation value in the vacuum of the CFT.

As noted in Ref.\cite{WFfromCB}, this construction is very similar to matrix product states (MPS) obtained from CFT \cite{iMPS, iMPS2}. In both cases, the ancillary degrees of freedom are described by a quantum field theory and the resulting many-body wave functions describes a lattice system. As a matter of fact, note that $\ket{\psi_{ee}}$ corresponds to the many-body state defined in that paper written in fermionic variables
\begin{equation}
\Psi_\mm^{(ee)} = \braket{\mm | \psi_{ee}} = \F_\mm (z_1,\cdots, z_{2n}),
\label{CBamplitudes}
\end{equation}
where $\ket{\mm} = \ket{m_1\cdots m_N}$. The present formulation highlights both the inner (i.e., entanglement) structure of these states and its relation to the physical degrees of freedom.

We can also contruct other states by adding fermions to the asymptotic states (within the operator-state correspondence \cite{diFrancesco}), in particular
\begin{align}
\ket{\psi_{oo}} = \braket{ \chi\left| \Phi_{oo}^{(N)} \right| \chi} \ket{0}_c \in\H_e.
\label{psioo}
\end{align}
As we will see in a later section, these wave functions are natural ans\"atze for low-energy excited eigenstates.


\section{First-order picture: OPE Analysis}

The construction we have discussed so far is quite general. In order to get a more intuitive picture of these states, we can consider a first-order approximation using the operator product expansion (OPE). This scheme will allow us to get a glimpse of the structure of state $\ket{\psi_{ee}}$ using simplified operators.

The full expression of the OPE of two $\sigma$ fields is given by \cite{diFrancesco}
\begin{align}
\sigma(z_1)\sigma(z_2) = \frac{1}{z_{12}^{1/8}}\bigg(&\sum_{\alpha\in\V_\id}z_{12}^{h_\alpha}C_{\sigma\sigma}^\alpha \alpha\left(\frac{z_1 + z_2}{2}\right)
\label{fullSigmaOPE}\\
&+\sum_{\beta\in\V_\chi}z_{12}^{h_\beta}C_{\sigma\sigma}^\beta \beta\left(\frac{z_1 + z_2}{2}\right)\bigg),\nonumber
\end{align}
where $z_{12} = z_1 - z_2$, $\alpha$ $(\beta)$ are the fields with conformal weights $h_\alpha$ $(h_\beta)$ that generate the Verma module $\V_\id$ $(\V_\chi)$ by acting on the vacuum, and $C_{\sigma\sigma}^\alpha, C_{\sigma\sigma}^\beta$ are constants fixed by 3-point functions. (Note that we are using a symmetrized version of the OPE, instead of pinning the resulting operators on $z_2$.) If we only keep the lowest orders in the expansion, we get the familiar expression
\begin{equation}
\sigma(z_1)\sigma(z_2) \sim \frac{1}{z_{12}^{1/8}}\left(1+\left(\frac{z_{12}}{2}\right)^{1/2}\chi\left(\frac{z_1+z_2}{2}\right)\right),
\label{OPEapprox}
\end{equation}
where we used the fact that $C_{\sigma\sigma}^\chi = 1/\sqrt{2}$.

Assume now that we have $N$ pairs of $\sigma$ fields, parametrized by
\begin{equation}
z_{2n-1} = w_n - \frac{1}{2}\delta_n, \qquad z_{2n} = w_n +\frac{1}{2} \delta_n.
\end{equation}
Using this notation and the OPE, we can write the approximate expression for \eqref{Amatrix}
\begin{equation}
A^{(n)} \sim \frac{1}{\delta_n^{1/8}}\left[\id_2 +\left( \sqrt{\frac{\delta_n}{2}}c_n^\dg\chi(w_n)\right)\sigma^x \right],
\label{AmatrixOPE}
\end{equation}
where $\sigma^x =   \begin{pmatrix}
 0 & 1 \\
   1 & 0
 \end{pmatrix}$ is one of the Pauli matrices. Note that this approximation implies
 \begin{equation}
 V_{\id \id}^\id = V_{\chi \chi}^\id \sim \frac{1}{\delta_n^{1/8}}\id,  \quad V_{\id\chi}^\chi = V_{\chi \id}^\chi \sim \frac{1}{\sqrt{2}}\delta_n^{3/8}\chi(w_n).
 \label{OPEvertex}
 \end{equation}

Given that $(c_n^\dg)^2=0$, we have
\begin{equation}
A^{(n)} \propto \exp\left(\sqrt\frac{\delta_n}{2} c_n^\dg \chi(w_n)\sigma^x\right),
\end{equation}
so that \eqref{PsiMatrix} becomes
\begin{equation}
\Phi^{(N)} \propto \exp\left[\sum_{n=1}^N\left(\sqrt\frac{\delta_n}{2} c_n^\dg \chi(w_n)\right)\sigma^x \right].
\end{equation}
Now, using the fact that the vacuum of the Ising CFT is a free Gaussian state for the Majorana fermions \cite{diFrancesco}, we employ the familiar identity
\begin{equation}
\braket{\exp(A)}_\text{Gaussian} = \exp\left(\frac{1}{2}\braket{A^2}_\text{Gaussian}\right)
\end{equation}
(assuming $\braket{A}_\text{Gaussian}=0$) to obtain
\begin{align}
\braket{0\left|\Phi_{ee}^{(N)}\right|0} &\propto \exp\left[\sum_{n<m}\frac{\sqrt{\delta_n\delta_m}}{2}\braket{\chi(w_n)\chi(w_m)}c_n^\dg c_m^\dg\right]\nonumber\\
&=\exp\left[\sum_{n<m}\frac{\sqrt{\delta_n\delta_m}}{2\left(w_n - w_m\right)}c_n^\dg c_m^\dg\right].
\end{align}
We conclude then that $\ket{\psi_{ee}}$ is a BCS state defined by the (real-space) pairing function
\begin{equation}
g_{nm}^{(\text{OPE})} = \frac{\sqrt{\delta_n\delta_m}}{2(w_{n}-w_{m})}.
\label{OPEpairingPlane}
\end{equation}
Note that this result holds for arbitrary complex coordinates and only depends on the validity of the OPE approximation. One may wonder if this expansion is really needed to guarantee the BCS structure of the lattice wave function. As we will show in a later section, numerical calculations suggests that this result extends beyond the OPE expansion, albeit with a different pairing function. We will also discuss some aspects regarding a full analytical proof of this fact.

If $|z_n|=1$, it is also convenient to use the conformal transformation that maps the plane to the cylinder
\begin{equation}
z \mapsto \exp(i\theta). 
\end{equation}
In this setting, we parametrize the coordinates as
\begin{equation}
\theta_{2n-1} = \phi_n - \frac{1}{2}\epsilon_n, \qquad \theta_{2n} = \phi_n + \frac{1}{2}\epsilon_n,
\label{cylcoord}
\end{equation}
so that the OPE can be written as
\begin{align}
\sigma(\theta_{2n-1})& \sigma(\theta_{2n}) \\
&\sim \left(\frac{1}{2\sin\left(\frac{\epsilon_n}{2}\right)}\right)^{1/8}\left(1+\sin^{1/2}\left(\frac{\epsilon_n}{2}\right)\chi(\phi_n)\right).\nonumber
\end{align}
Given that on the cylinder we have
\begin{equation}
\braket{\chi(\phi_1)\chi(\phi_2)}_{\text{cyl}} = \frac{1}{2\sin\left(\frac{\phi_1-\phi_2}{2}\right)},
\end{equation}
a similar analysis yields a BCS state with the pairing function
\begin{equation}
g_{nm}^{(\text{OPE, cyl})} = \frac{\sqrt{\sin\left(\frac{\epsilon_n}{2}\right)\sin\left(\frac{\epsilon_m}{2}\right)}}{2\sin\left(\frac{\phi_n - \phi_m}{2}\right)}.
\label{OPEpairingCylinder}
\end{equation}
This representation is particularly useful for lattice configurations which are periodic, such as cylinders. Note again that this analysis holds for arbitrary configurations, allowing for complex $\theta_n$.


\section{Exact expressions for the Ising conformal blocks}

Finding the exact form of the CBs for an arbitrary CFT is in general an ardous task. While it is known that they must satisfy a set of well-known differential equations \cite{diFrancesco}, it is far from obvious that they can be solved analytically for any given number of primary fields. In the case of the Ising CFT, one can find exact closed expressions by means of bosonization \cite{NayakWilczek, IsingCB}.

We will make use of the multiperipheral basis to write the exact formulas for the CBs. This is a canonical representation that is valid for all types of CBs \cite{QuantumGroupsCFT,WFfromCB}. We will omit the $\sigma$'s in this notation and write $\pp=(p_1,\cdots,p_{N-1})$, where $p_i=0$ ($1$)  corresponds to an identity operator $\id$ (a fermion $\chi$).  (Note that $\pp$ can take $2^{N-1}$ different values, as expected.)

\begin{figure}[h]
  \centering

\includegraphics[width=0.95\linewidth]{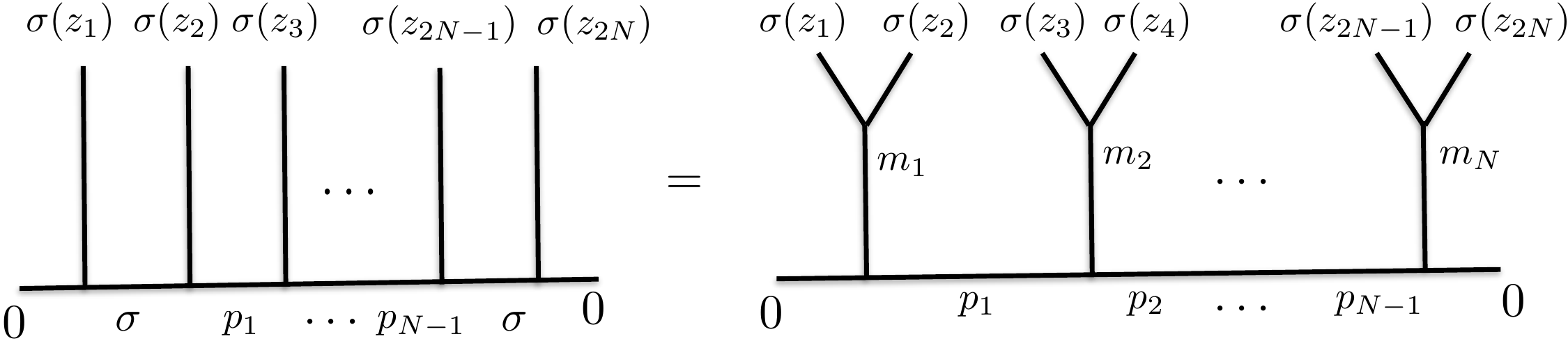}

    \caption{A conformal block using only $\si$ field operators grouped in reference pairs $(\si(z_{2k-1}), \si(z_{2k}))$. The equivalance between the two representations is obtained from the relation $m_k = p_{k-1} + p_k (\text{mod}\, 2)$.}
    \label{IsingCB}
\end{figure}

We can easily relate the multiperipheral basis to the one obtained from the pair-wise fusion of operators (see Fig. \eqref{IsingCB}). The latter is the basis that we used previously in \eqref{CBamplitudes}. Note that, in order to preserve the number of fermions at each vertex, there is the restriction $m_k = p_{k-1} + p_k (\text{mod}\, 2)$. (We define fixed auxiliary values $p_0=p_N=0$.)

Before stating the formulas for the CBs, we introduce some extra notation. First, we will need certain bipartitions of the $\si$ field coordinates that associate the points of each reference pairs to different groups. We call these macrogroups $\ell_\qq, \ell'_\qq$ and they are generated from an integer $\qq=0,\cdots, 2^{N-1}-1$ according to \cite{WFfromCB, IsingCB}
%
%
%
\begin{equation}
\ell_\qq(k) = 2k -\frac{1}{2}(1+s_k), \qquad \ell'_\qq(k) = 2k -\frac{1}{2}(1-s_k),
\label{Spin macrogroup}
\end{equation}
where $q_k$ are the binary digits of $\qq = (q_1, q_2, \dots, q_{N-1})$,
\begin{equation}
s_k = \prod_{i=1}^{k-1}\left(1-2q_i\right),
\label{auxspin}
\end{equation}
and $s_1=1$ by definition.

Using this notation, we can define
\begin{equation}
z_{\ell_\qq}=\prod_{k<m}z_{\ell_\qq(k),\ell_\qq(m)}.
\end{equation}
where $z_{ab} = z_a - z_b$. We will also need the sign given by
\begin{align}
\epsilon_{\pp\qq} &\equiv (-1)^{\sum_k p_k q_k}= \prod_{k=1}^{N-1}\left(1-2p_k q_k\right)\\
&= \prod_{k=1}^{N-1}\left(1+p_k (s_{k}s_{k+1}-1)\right)\equiv \tilde\epsilon_{\pp\sss}.\nonumber
\end{align}
using the binary expansion of both $\pp$ and $\qq$.

The expression for the CB can be written as \cite{IsingCB}
\begin{equation}
\F^{(2N)}_\pp = \frac{1}{2^{\frac{N-1}{2}}}\prod_{a<b}^{2N}z_{ab}^{-1/8}\left(\sum_{\qq = 0}^{2^{N-1}-1}\ep_{\pp\qq}\sqrt{z_{\ell_\qq}z_{\ell'_\qq}}\right)^{1/2}.
\label{CBsigmas}
\end{equation}
Note that the sum inside the square root is the only part that depends on $\pp$. 

It is important to remark that we are assuming radial ordering
\begin{equation}
|z_1|\geq |z_2| \geq \cdots \geq |z_{2N}|. 
\end{equation}
Moreover, if  $|z_n| = |z_m|$ and $n<m$, we will assume that the angular parts in the polar decomposition are ordered with respect to the principal value of the logarithm. In other words, if $z_n = \exp(a_n + ib_n)$, whenever $a_n = a_m$, we will assume
\begin{equation}
-\pi < b_n < b_m \leq \pi
\end{equation}
if $n<m$.

The ordering of the coordinates will be important because it ensures that we consistently choose the same branches of the (complex) square root. In order to see this, let us define
\begin{equation}
B_\qq =  \prod_{n<m}^N\left[\left(1 - \frac{z_{\ell_\qq(m)}}{z_{\ell_\qq(n)}}\right)\left(1 - \frac{z_{\ell'_\qq(m)}}{z_{\ell'_\qq(n)}}\right)\right]^{\frac{1}{2}}.
\end{equation}
Using this notation, we note that we can write the $\pp$-dependent part of \eqref{CBsigmas} using only the main branch of the square root
\begin{equation}
\F_\pp^{(2N)} \propto \left(\sum_{\qq = 0}^{2^{N-1}-1}\ep_{\pp\qq} \frac{B_\qq}{B_0}\right)^{1/2}.
\end{equation}
This will be particularly important for 2D spin configurations. Note that we can obtain CBs for coordinates which are not radially ordered by analytic continuation of these expressions. This can be done by means of the Ising braid matrices \cite{QuantumGroupsCFT,WFfromCB, MooreSeibergNotes}.


\section{1D wave functions}

\begin{figure}[h]
  \centering

\includegraphics[width=0.65\linewidth]{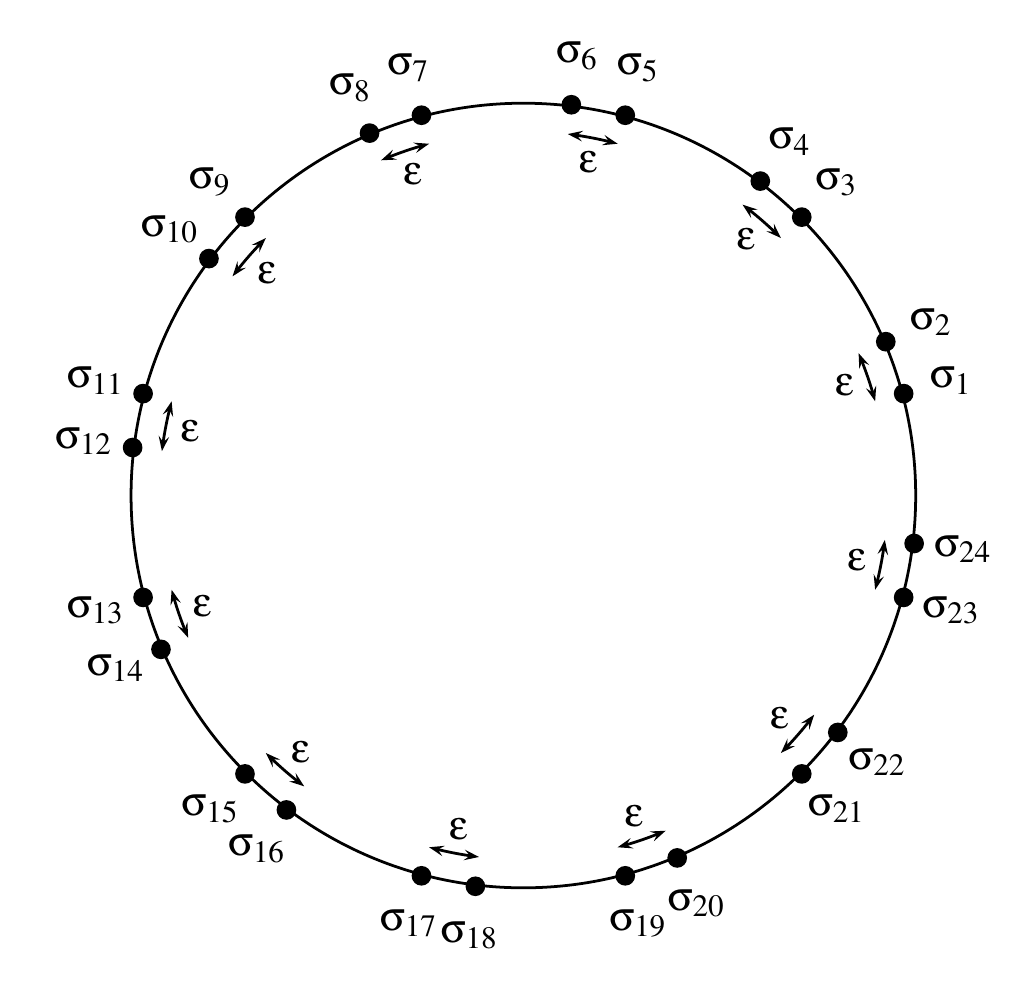}

    \caption{Coordinate configuration on the complex plane for a 1D system with 12 spins. Note that the reference pairs are uniformly distributed, so the wave function is translationally invariant for all values of $\epsilon$.}
    \label{conf_epsilon}
\end{figure}

We focus now on a one-dimensional configuration. For this purpose, we choose the $2N$ coordinates to be given by $z_k=\exp(i\theta_k)$, where (see Fig. \eqref{conf_epsilon})
\begin{equation}
\theta_k =\frac{2\pi}{2N}\left(k+(-1)^k\delta-N\right),
\label{coord1d}
\end{equation}
and $\delta\in (-\frac{1}{2},\frac{1}{2})$ is a fixed parameter. Using this parametrization, we can rewrite the wavefunction amplitudes \eqref{CBamplitudes} as \cite{WFfromCB}
\begin{equation}
\Psi_\pp^{(ee)}(\delta) = \frac{1}{\tilde N_0}\left(\sum_{\qq = 0}^{2^{N-1}-1}\ep_{\pp\qq}\, A_\qq(\delta)\right)^{1/2},
\label{varWF}
\end{equation}
where
\begin{equation}
A_\qq = \prod_{n>m}^N\left[\sin\frac{\theta_{\ell_\qq(n)}-\theta_{\ell_\qq(m)}}{2}\sin\frac{\theta_{\ell'_\qq(n)}-\theta_{\ell'_\qq(m)}}{2}\right]^{\frac{1}{2}},
\label{AAA}
\end{equation}
and $\tilde N_0$ is a normalization constant
\begin{equation}
\tilde N_0^2 = \frac{N^{N/2}}{2^{(N-1)(N-2)/2}}.
\end{equation}

We can also write \eqref{AAA} in terms of the auxiliary spins \eqref{auxspin} (see the Appendix in Ref.\cite{WFfromCB})
\begin{equation}
A(\{s_k\}) = \prod_{j>i}^N \sin\left[\frac{\pi}{N}\left(j-i+\frac{1+2\delta}{4}(s_j-s_i)\right)\right].
\label{AAAspins}
\end{equation}
Note that for all values of $\delta$, the resulting wave function describes a translationally invariant spin chain with periodic boundary conditions. This is a consequence of the fact that we are describing physical degrees of freedom on the lattice by means of pairs of $\sigma$ fields. The centers-of-mass of the pairs are uniformly distributed on the circle, while their size is constant for fixed $\delta$
\begin{align}
\theta_{2k}-\theta_{2k-1} &= \frac{2\pi}{N}(\frac{1}{2}+\delta)\equiv\frac{2\pi}{N}\epsilon, \\
\frac{\theta_{2k}+\theta_{2k-1}}{2} &= \frac{2\pi}{N}\left(k-\frac{2N+1}{2}\right).\nonumber
\end{align}

In this representation, there is an exponentially large number of numerical operations that need to be performed. Luckily, we can obtain a determinant form for these particular configurations that simplifies the calculations. Once again, we make use of the pair-wise fusion basis $\mm=(m_1,\cdots,m_N)$. It can be shown that the normalized wave function amplitudes can be written as (we will leave the details for Appendix A)
\begin{equation}
\Psi^{(ee)}_\mm(\delta) = \left(\frac{\det(F_\mm*V)}{\det(V)}\right)^{1/2},
\label{psieedet}
\end{equation}
where $F_\mm*V$ is the element-wise matrix product (also known as the Hadamard product of matrices)
\begin{equation}
(F_\mm*V)_{rt} = (F_\mm)_{rt}(V)_{rt},
\end{equation}
obtained from matrices
\begin{equation}
(V)_{rt} = \exp\left(i\frac{2\pi}{N}r(t-1)\right)
\end{equation}
and
\begin{equation}
(F_\mm)_{rt} = \left\{
  \begin{array}{l l}
    \cos\left[\frac{\pi}{4N}(1+2\delta)\left(2t-N-1\right)\right], & \,\, m_r=0,\\
    i\sin\left[\frac{\pi}{4N}(1+2\delta)\left(2t-N-1\right)\right], & \,\, m_r=1.
  \end{array} \right. 
\end{equation}

Determinant expression \eqref{psieedet} can also allow us to write $\ket{\psi_{ee}(\delta)}$ as a BCS state. If we define the lattice momenta as
\begin{equation}
k = \frac{\pi}{N}(2m-N-1), \qquad m=1,\cdots,N
\label{kdef}
\end{equation}
we can write the normalized state as
\begin{equation}
\ket{\psi_{ee}(\delta)} = C_N(\delta) \exp\left(\sum_{n<m}\tilde g_{nm}(\delta)c_n^\dg c_m^\dg \right)\ket{0}_c,
\end{equation}
where
\begin{equation}
C_N(\delta) = \prod_k \sqrt{\cos\left[\frac{(1+2\delta)}{4}k\right]},
\label{cN1D}
\end{equation}
%
and $\tilde g_{nm}=g_{n-m}$ with
\begin{align}
g_r = (-1)^r\frac{2}{N}\sum_{k>0} & \tan\left[\frac{(1+2\delta)}{4}k\right]\sin\left(kr\right).
\label{g!D}
\end{align}
%
We will leave the details for Appendix B. Note that this result is exact and does not depend on any approximation. We also highlight that this is further evidence that $\ket{\psi_{ee}}$ as defined in \eqref{psiee} has a BCS structure beyond the OPE regime.


\section{The ground state of the critical Ising spin chain from conformal blocks}

In Ref.\cite{WFfromCB}, it was argued from numerical evidence that $\ket{\psi_{ee}(\delta=0)}$ correspond exactly to the ground-state of the even sector (defined by $\braket{Q}=\braket{\prod_n\sigma^z_n}=1$) of the Ising tranverse field (ITF) critical Hamiltonian with periodic boundary conditions
\begin{equation}
H = -\sum_{n=1}^N \sigma^x_n\sigma^x_{n+1} - \sum_{n=1}^N \sigma^z_n.
\label{ITFh1}
\end{equation}
We will now present a analytical proof of this result. 

The exact solution of \eqref{ITFh1} is well-known \cite{Henkel, Sachdev, Mussardo}. The ground state can be obtained by mapping the spin variables to spinless fermions using a Jordan-Wigner (JW) transformation
\begin{align}
\sigma_n^z &= 1-2c_n^\dg c_n, \\ 
\sigma_n^x &= \prod_{m=1}^{n-1}(1-2c_m^\dg c_m)(c_m^\dg+c_m),\nonumber
\end{align}
to obtain
\begin{align}
H = -\sum_{n}(1-2c_n^\dg c_n)- \sum_{n}(c_n^\dg - c_n)(c_{n+r}^\dg + c_{n+r}).
\end{align}
This is a translationally invariant quadratic Hamiltonian that can be solved via a Fourier transform
\begin{equation}
c_n^\dg = \frac{1}{\sqrt{N}}\sum_k e^{ikn}c_k^\dg,
\end{equation}
%
where we take $k$ as in \eqref{kdef}, followed by a Bogoliubov transformation. The normalized ground state has a BCS structure
\begin{align}
\ket{gs} &=  \prod_{k>0} \left[\cos\left(\frac{\theta_k}{2}\right)+i\sin\left(\frac{\theta_k}{2}\right)c_k^\dg c_{-k}^\dg\right]\ket{0}_c \\
&=\prod_{k>0} \cos\left(\frac{\theta_k}{2}\right)\exp\left[\sum_{k>0}i\tan\left(\frac{\theta_k}{2}\right)c_k^\dg c_{-k}^\dg\right]\ket{0}_c,\nonumber
\end{align}
where we define
\begin{align}
\cos\left(\frac{\theta_k}{2}\right) &=\sqrt{\frac{1+\sin\left|\frac{k}{2}\right|}{2}},\\
\sin\left(\frac{\theta_k}{2}\right) &= -\text{sgn}(k)\sqrt{\frac{1-\sin\left|\frac{k}{2}\right|}{2}}. \nonumber
\end{align}
This implies that the normalization constant is
\begin{align}
\prod_{k>0} \cos\left(\frac{\theta_k}{2}\right) &= \prod_{k>0}\sqrt{\frac{1+\sin\left(\frac{k}{2}\right)}{2}}\\
&=\prod_{m=1}^N \sqrt{\cos\left[\frac{\pi}{4N}\left(2m-N-1\right)\right]}. \nonumber
\end{align}
We can also compute the real-space pairing function by doing a Fourier transform of $g_k=i\tan\left(\frac{\theta_k}{2}\right)$
%
%
\begin{align}
g_{r}&=\frac{2}{N}\sum_{k>0}\frac{1-\sin\left(\frac{k}{2}\right)}{\cos\left(\frac{k}{2}\right)}\sin\left(kr\right)\\
&=(-1)^r\frac{2}{N}\sum_{k>0}\tan\left(\frac{k}{4}\right)\sin\left(kr\right),\quad r\in\mathbb{Z},\nonumber
\end{align}
where the second expression is obtained from the first one by replacing $k\mapsto \pi-k$. 

Now, coming back to the wave functions obtained from the Ising CBs, note that these expressions correspond to the normalization constant \eqref{cN1D} and the pairing function \eqref{g!D} obtained in the previous section when $\delta=0$, so that
\begin{equation}
\ket{gs}=\ket{\psi_{ee}(\delta=0)}.
\end{equation}
This is a remarkable result given that the expression for the CBs \eqref{CBsigmas} was obtained from the infrarred fixed point of the critical theory. It is non-trivial that it would agree with the ground state of a finite-size lattice system.


\section{Parent Hamiltonians for 1D}

We have checked numerically that for $\delta\neq 0$, we can find parent Hamiltonians that can also be mapped to a quadratic fermionic form. We consider the following family of Hamiltonian terms
\begin{align}
Z &= -\sum_{n}\sigma_n^z, \nonumber\\
X_r &=  -\sum_n \sigma_{n}^x\sigma_{n+1}^z\cdots\sigma_{n+r-1}^z\sigma_{n+r}^x,\label{HamTerms1D} \\
Y_r &= -\sum_n \sigma_{n}^y\sigma_{n+1}^z\cdots\sigma_{n+r-1}^z\sigma_{n+r}^y,\nonumber
\end{align}
with $r=1,\cdots,N/2$. (Note that $X_1$ is the usual Ising term.) This particular choice for the family of Hamiltonian terms corresponds to those that will yield quadratic forms in fermionic variables (see Appendix C for an explicit fermionic formulation). Given that $\ket{\psi_{ee}(\delta)}$ is translationally invariant and describes a system with periodic boundary conditions, we impose the same constraints on the Hamiltonian terms. 

We know that the variational wavefunctions obtained from the CB of the Ising model behave nicely under a Kramers-Wannier (KW) duality transformation  \cite{WFfromCB, topDefIsing}. In particular, we have that
\begin{equation}
\ket{\psi_{ee}(\delta)} \mapsto \ket{\psi_{ee}(-\delta)}.
\end{equation}
Something similar can be said about the Hamiltonian terms we are considering.

\begin{figure}[h]
  \centering

\includegraphics[width=1.0\linewidth]{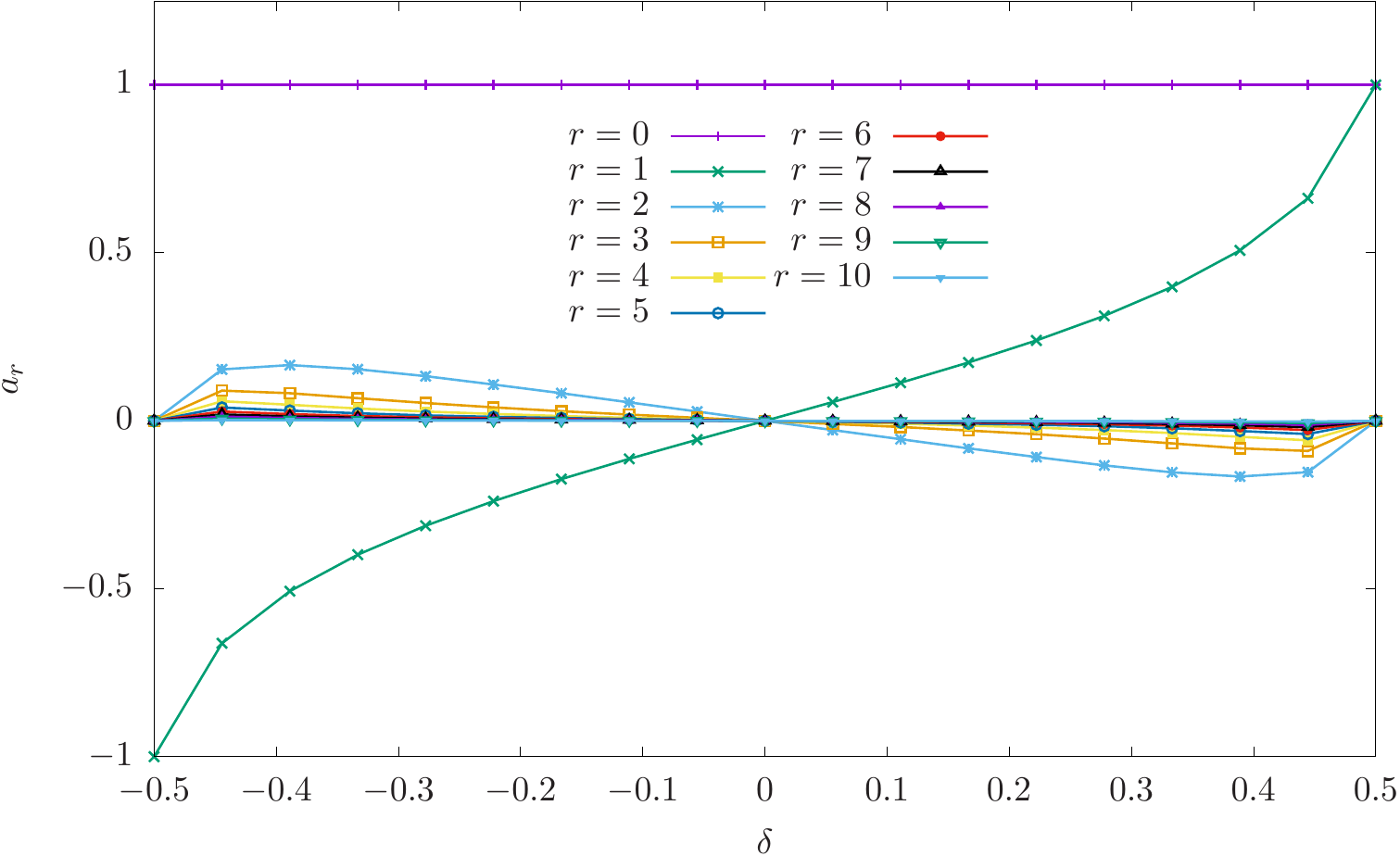}

    \caption{Variational coefficients of the parent Hamiltonians of the form \eqref{varHams} for $\ket{\psi_{ee}(\delta)}$ for $N=20$ spins. They are normalized so that $a_1=1$. }
    \label{coefham}
\end{figure}

The action of the KW transformation can be summarized in the map
\begin{equation}
\sigma_n^z \mapsto \sigma_n^x \sigma_{n+1}^x, \qquad \sigma_n^x \mapsto \sigma_1^z\cdots \sigma_n^z.
\end{equation}
(See Appendix C for a formulation of the KW transformation in terms of Majorana fermions.)From these relations, it is easy to compute the action of the KW transformation (for the even-parity sector of the Hilbert space, defined by $\braket{Q}=\braket{\prod_n\sigma^z_n}=1$)
\begin{align}
Z \mapsto X_1, \quad & X_1 \mapsto Z, \quad X_r \mapsto -Y_{r-1},\, (r=2,\cdots N/2), \nonumber\\ & Y_r \mapsto -X_{r+1},\, (r=1,\cdots, N/2).
\end{align}
Note first that the KW dual of a Hamiltonian that can be written as a quadratic form in fermionic variables is once again of the same type.

We can try to take advantage of the KW duality. For our variational fits, we used the Hamiltonian family
\begin{align}
&\tilde H_0 = \id, \quad \tilde H_1 = X_1 + Z, \quad \tilde H_2 = X_1 - Z,\label{HamFamily1D} \\
&\tilde H_r = X_{r-1} + Y_{r-2},\, (r=3,\cdots N/2+1).\nonumber
\end{align}
We need $\tilde H_0$ to be equal to the identity for the variational algorithm (see Appendix C). Notice also that $\tilde H_1$ corresponds to the critical ITF Hamiltonian \eqref{ITFh1}. Using these definitions, the whole family is closed under a KW transformation
\begin{align}
&\tilde H_0 \mapsto \tilde H_0, \quad \tilde H_1 \mapsto \tilde H_1, \\
&\tilde H_r \mapsto -\tilde H_r, \, (r=2,\cdots, N/2+1).\nonumber
\end{align}

We checked numerically that the wavefunction $\ket{\psi_{ee}(\delta)}$ obtained from the Ising CBs with $|\delta|<1/2$ is the ground state of a Hamiltonian of the form (see Fig. \eqref{coefham})
\begin{equation}
H =  -\sum_{r=1}^{N/2+1}a_r \tilde H_r.
\label{varHams}
\end{equation}
The duality implies that $a_1(\delta) = a_1(-\delta)$, so we will set $a_1=1$. (Remember this is the coefficient associated to the critical ITF Hamiltonian term.) Note also that
\begin{equation}
a_r(-\delta) = -a_r(\delta), \,(r=2,\cdots,N/2+1).
\end{equation}

In Fig. \eqref{coefham}, we plot the variational coefficients obtained for different values of $\delta$ and $N=20$ spins. We see that the Hamiltonian is dominated by $\tilde H_1$, namely, the ITF critical Hamiltonian \eqref{ITFh1}. The other significant contribution comes from $\tilde H_2$, in particular for $|\delta|\approx 1/2$, so that the ground state approximates the trivial Ising fixed points in these limits (see Ref.\cite{WFfromCB}).

For $|\delta|\approx 0$, all the Hamiltonian terms that change sign under a KW transformation are very small compared to $\tilde H_1$. Given that the whole Hamiltonian family $\{\tilde H_r\}$ respect the basic Ising symmetries, this implies that $\ket{\psi_{ee}(\delta)}$ approximates small massive perturbations away from criticality. If we drop the Hamiltonian terms for $r>2$, we see that in this vicinity the corresponding transverse field will be given by
\begin{equation}
h = \frac{1-a_2}{1+a_2}\approx 1-2a_2 + \mathcal{O}(a_2^2).
\end{equation}
This explains the relative good agreement between $\ket{\psi_{ee}(\delta)}$ for $|\delta|\ll 1$ and the ground state of the ITF Hamiltonian close to the critical point \cite{WFfromCB}.


\section{Excited states}

We can extend the previous discussion to other states obtained from operator matrix \eqref{PsiMatrix}. Let us first consider state $\ket{\psi_{oo}}$, defined in \eqref{psioo}. In this case, the asymptotic states of the CFT are fermions
\begin{equation}
\Psi^{(oo)}_\pp \propto \braket{\chi | \sigma(z_1)\cdots \sigma(z_{2N}) | \chi}_\pp.
\end{equation}
Assuming radial ordering, we can obtain the amplitudes for this state by adding two fermions at $z=0,\infty$. Starting from the exact expression  \cite{IsingCB} and taking the appropriate limit, the corresponding amplitudes for the associated wave function are given by
\begin{align}
\Psi^{(oo)}_\pp &= \frac{1}{\tilde N_2}\left(\sum_{\qq = 0}^{2^{N-1}-1}\ep_{\pp\qq}A_\qq\right)^{-1/2}\\
&\left[\sum_{\qq = 0}^{2^{N-1}-1}\ep_{\pp\qq}A_\qq\left( \sqrt{\prod_{k=1}^N\frac{z_{\ell_\qq(k)}}{z_{\ell'_\qq(k)}}} + \sqrt{\prod_{k=1}^N\frac{z_{\ell'_\qq(k)}}{z_{\ell_\qq(k)}}} \right)\right].\nonumber
\end{align}
If we use the homogeneous 1D configuration \eqref{coord1d}, we can rewrite these amplitudes as
\begin{align}
\Psi^{(oo)}_\pp &= \frac{1}{N_2}\left(\sum_{\{s_k\}}\tilde\epsilon_{\pp\sss}\,A(\{s_k\})\right)^{-1/2}\\
&\left(\sum_{\{s_k\}}\tilde\epsilon_{\pp\sss}\,A(\{s_k\})\cos\left[\frac{\pi}{2N}(1+2\delta)\sum_{k=1}^N s_k\right]\right.,\nonumber
\end{align}

They can also be written in terms of determinants. Define the matrix
\begin{equation}
(J_\mm(q))_{r,t} = \exp\left(i\frac{2\pi}{N}r(t-1)\right)(F_\mm(q))_{r,t}\, ,
\end{equation}
where
\begin{align}
(F_\mm(q)&)_{r,t} \\
=&\left\{
  \begin{array}{l l}
    \cos\left[\frac{\pi}{4N}(1+2\delta)\left(2t-N-1+q\right)\right], & \, m_r=0,\\
    i\sin\left[\frac{\pi}{4N}(1+2\delta)\left(2t-N-1+q\right)\right], & \,m_r=1.
  \end{array} \right. \nonumber
\end{align}
We can easily show that
\begin{equation}
\Psi^{(oo)}_\mm \propto \frac{1}{\left(\det(J_\mm(0)\right)^{1/2}}\left[\det(J_\mm(2)+\det(J_\mm(-2)\right].
\end{equation}

Given that this wave function can be written in terms of real amplitudes (up to a possible overall phase that does not depend on $\pp$), it is easy to compute the overlap with  $\ket{\psi_{ee}}$
\begin{align}
\braket{\psi_{ee}(\delta)|\psi_{oo}(\delta)}&=\sum_\pp \Psi^{(ee)}_\pp \Psi^{(oo)}_\pp \\
&\propto \cos\left(\frac{\pi}{2}(1+2\delta)\right) = -\sin\left(\pi\delta\right).\nonumber
\end{align}
This implies that the two states will be orthogonal if $\delta=0$. (Recall that we are assuming $|\delta|<1/2$.) This is exactly the case for which $\ket{\psi_{ee}}$ describes the ground state of the critical ITF Hamiltonian \eqref{ITFh1}.

We have checked numerically the action of this Hamiltonian on $\ket{\psi_{oo}(\delta=0)}$ for sizes up to $N=20$ spins using a Lanczos algorithm. We found that it corresponds within machine precision to the first excited state of the even-parity sector of the critical ITF Hamiltonian \eqref{ITFh1}. This is again a remarkable result given that the amplitudes are computed from CBs obtained at the infrared limit of the thermodynamic model.

These results reflect the relation between the finite-size study of the Ising spin chain and the operator content of the Ising CFT \cite{Henkel, ConformallyInvariantLimit, CardyContent}. It is known that the spectrum of the even-parity sector of \eqref{ITFh1} with periodic boundary conditions corresponds to the Virasoro towers of both operators $\id$ and $\epsilon$. These are the primary operators of the full Ising CFT that are even under the internal $\mathbb{Z}_2$ symmetry. Starting from these states, we can in principle construct the full spectrum of the Hamiltonian by acting with the corresponding representation of the Virasoro algebra. These operators can be obtained on the lattice from the local Hamiltonian density \cite{SKoriginal, SKnonintegrable}.

It is tempting to extend this construction to the odd-parity sector of the ITF spin chain. Finite-size scaling using periodic boundary conditions relate this sector to the Virasoro tower of $\sigma$ \cite{Henkel}. We tried the natural candidates obtained from (a) using a single fermion on the asymptotic states, both at $z=0,\infty$, so that the CFT degrees of freedom are traced out by $\bra{0}\cdots\ket{\chi}$ or $\bra{\chi}\cdots\ket{0}$; (b) using a pair of $\sigma$ fields on the asymptotic states $\bra{\sigma}\cdots\ket{\sigma}$. In both scenarios, the amplitudes obtained using configuration \eqref{coord1d} for $\delta=0$ contained complex amplitudes that cannot be factored to an overall phase. This implies that these states cannot be used naively to describe ground states of real Hamiltonians. Moreover, using $\sigma$ for both asymptotic states can yield wave functions that are not translationally invariant even if the degrees of freedom are arranged uniformly on the circle. This suggests that there is a richer structure underlying the general framework that needs to be understood in further work. 


\section{BCS structure beyond OPE: general observations}

We have seen that the OPE expansion of the CBs yields many-body wave functions with a BCS structure, and that this remains true in the exact case for translationally invariant 1D configurations. One may wonder if this result still holds true for the exact CBs using an arbitrary configuration (assuming, of course, radial ordering). In order to check this, let us consider $N=4$ spins described by $\ket{\psi_{ee}}$. This provides the smallest system size in which a BCS wave function is non-trivial and it allows us to understand the problem in more detail.

First, consider a general BCS wave function for $N=4$. If we write it in full detail, we have
\begin{equation}
\ket{\psi_\text{BCS}} \propto \left(1 + \sum_{n<m}g_{nm}c_n^\dg c_m^\dg + g_{1234}c_1^\dg c_2^\dg c_3^\dg c_4^\dg\right)\ket{0}_c,
\end{equation}
where we define for convenience
\begin{equation}
g_{1234} = g_{12} g_{34} - g_{13} g_{24} + g_{14}g_{23} .
\label{BCS4}
\end{equation}
Note that this definition relates explicitly to Wick theorem for fermions. If $\ket{\psi_{ee}}$ does indeed describe a BCS state, we expect its amplitudes to fulfill this constraint.

\begin{figure}[h]
  \centering

\includegraphics[width=1.0\linewidth]{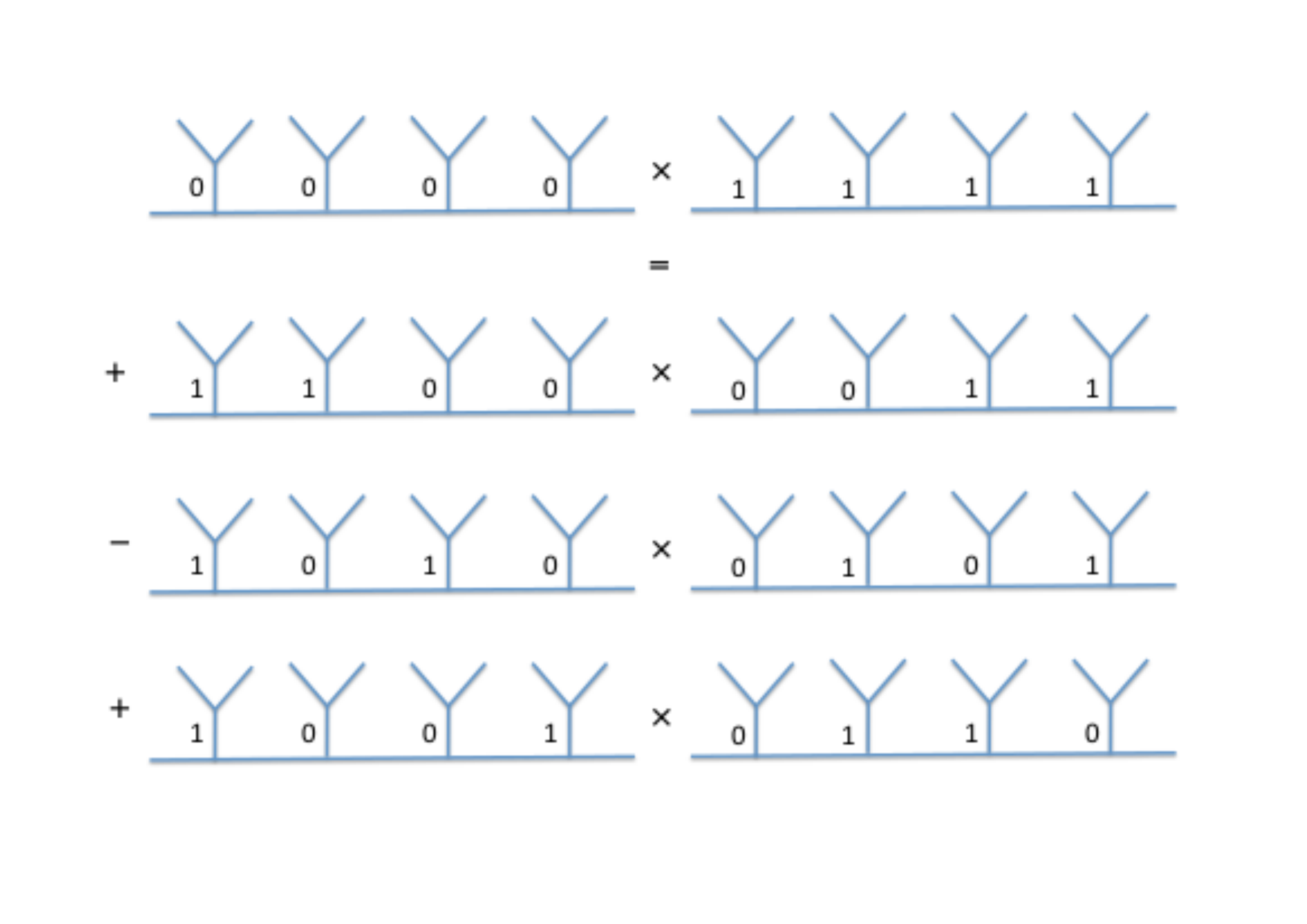}

    \caption{Graphical representation of equation \eqref{N4VertexCondition}. }
    \label{N4Vertex}
\end{figure}

In order to check this, let us write the operator matrix \eqref{Amatrix} as (we omit the coordinates for simplicity)
\begin{equation}
A^{(n)} = 
 \begin{pmatrix}
  V_{00} & c_n^\dg V_{01} \\
  c_n^\dg V_{10} & V_{11} 
 \end{pmatrix}.
\end{equation}
Using this notation, it is easy to see that condition \eqref{BCS4} will be fulfilled for $\ket{\psi_{ee}}$ if and only if (see Fig. \eqref{N4Vertex})
\begin{align}
\braket{V_{00}V_{00}V_{00}V_{00}}&\braket{V_{01}V_{10}V_{01}V_{10}}=\nonumber\\
&\braket{V_{01}V_{10}V_{00}V_{00}}\braket{V_{00}V_{00}V_{01}V_{10}}
\label{N4VertexCondition} \\
&- \braket{V_{01}V_{11}V_{10}V_{00}}\braket{V_{00}V_{01}V_{11}V_{10}}\nonumber \\
&+\braket{V_{01}V_{11}V_{11}V_{10}}\braket{V_{00}V_{01}V_{10}V_{00}}.\nonumber
\end{align}
Note first that this equation is trivially satisfied if all $V_{ij}$ are numbers. Also, if we use the OPE approximation \eqref{OPEvertex}, the equation reduces to the usual Wick theorem for free fermions. If we write this using the exact amplitudes in the pair-wise fusion basis, we get
\begin{equation}
\F_{0000}\F_{1111} = \F_{1100}\F_{0011} - \F_{1010}\F_{0101} + \F_{1001}\F_{0110}.
\end{equation}

We have checked numerically the condition for (radially ordered) random configurations using the exact CBs and they do indeed describe BCS wave functions. Unfortunately, we cannot provide a general proof even for such a small system size. One possible route is to expand the vertex operators using the full OPE expansion \eqref{fullSigmaOPE}. In that case, condition \eqref{N4VertexCondition} can be recast into a perturbative expression. Some subtleties regarding this approach are discussed in the Appendix D.


\section{2D wave functions}

So far, we have used coordinate configurations for the $\sigma$ fields that are constrained to the unit circle on the complex plane. We now study 2D configurations, where the formalism for the construction of the wave function will be very similar. Unfortunately, we cannot use the same procedure we described in Appendix B to write the amplitudes using a determinant form such as \eqref{psieedet}. This limits the system sizes we can consider numerically. However, we can get around this impasse by considering the OPE approximation we already discussed.

We will relate $\ket{\psi_{ee}}$ for a 2D configuration to the weak pairing phase of the effective mean-field Hamiltonian that describes $p+ip$ superconductivity \cite{ReadGreen}
\begin{equation}
H = \sum_{\kk} \left[\xi_\kk c_\kk^\dg c_\kk + \frac{1}{2}\left(\Delta_\kk^* c_{-\kk}c_\kk + h.c.\right) \right],
\label{HpSC}
\end{equation}
where
\begin{equation}
\xi_\kk =\frac{1}{2m}\kk^2 - \mu, \qquad \Delta_\kk = \hat\Delta(k_x - ik_y),
\end{equation}
$\mu$ is the chemical potential, and $\hat\Delta$ is a constant defining the gap function. The (normalized) ground state of this theory is obtained by usual BCS methods and can be written as
\begin{equation}
\ket{gs} = \left.\prod_\kk \right.' \left(u_\kk + v_\kk c_\kk^\dg c_{-\kk}^\dg\right)\ket{0},
\end{equation}
where the prime on the product indicates that each pair $(\kk,-\kk)$ appears only once, and $u_\kk,v_\kk$ are the Bogoliubov functions obtained from the Bogoliubov-de Gennes (BdG) equations
\begin{align}
E_\kk u_\kk = \xi_\kk u_k - \Delta_\kk^* v_\kk, \quad E_\kk v_\kk = -\xi_\kk v_\kk - \Delta_\kk u_\kk.
\end{align}
This reduces to
\begin{align}
E_\kk& = \sqrt{\xi_\kk^2 + |\Delta_\kk|^2},\\
|u_\kk|^2 &= \frac{1}{2}\left(1+\frac{\xi_\kk}{E_\kk}\right), \quad
 |v_\kk|^2 = \frac{1}{2}\left(1-\frac{\xi_\kk}{E_\kk}\right).\nonumber
\end{align}
The ground state can then be rewritten as
\begin{equation}
\ket{gs} =\left( \prod_\kk |u_\kk|^2\right)\exp\left(\frac{1}{2}\sum_\kk g_\kk c_\kk^\dg c_{-\kk}^\dg\right)\ket{0},
\end{equation}
where
\begin{equation}
g_\kk = \frac{v_\kk}{u_\kk} = -\frac{E_\kk-\xi_\kk}{\Delta_\kk^*}.
\end{equation}
(Note there is no restriction on $\kk$, except maybe for $\kk=0$.) Using the fermionic statistics, the amplitudes of the ground state can be written as Pfaffians \eqref{Pfaffian} using the real-space pairing function
\begin{equation}
g(\rr) = \frac{1}{N^2}\sum_\kk e^{i\kk\cdot\rr}g_\kk.
\end{equation}

If $\mu>0$, the system will be in the so-called weak pairing phase \cite{ReadGreen, Miguel}. For small momenta we have $\xi_\kk<0$ and
\begin{equation}
g_\kk \sim -\frac{2\mu}{\hat\Delta(k_x+ik_y)}.
\end{equation}
The leading behavior of the real-space pairing function is given by (see Appendix E for details)
\begin{equation}
g(\rr)\sim -\frac{2a^2 \mu}{2\pi i \hat\Delta}\frac{1}{x+iy},
\label{gSC}
\end{equation}
where $a$ is the lattice spacing. Note that this analysis is done on a regular square lattice, assuming a very large system size. However, the leading singular term gives the qualitative infrared behavior that determines the phase of the system.

\begin{figure}[h]
  \centering

\includegraphics[width=0.75\linewidth]{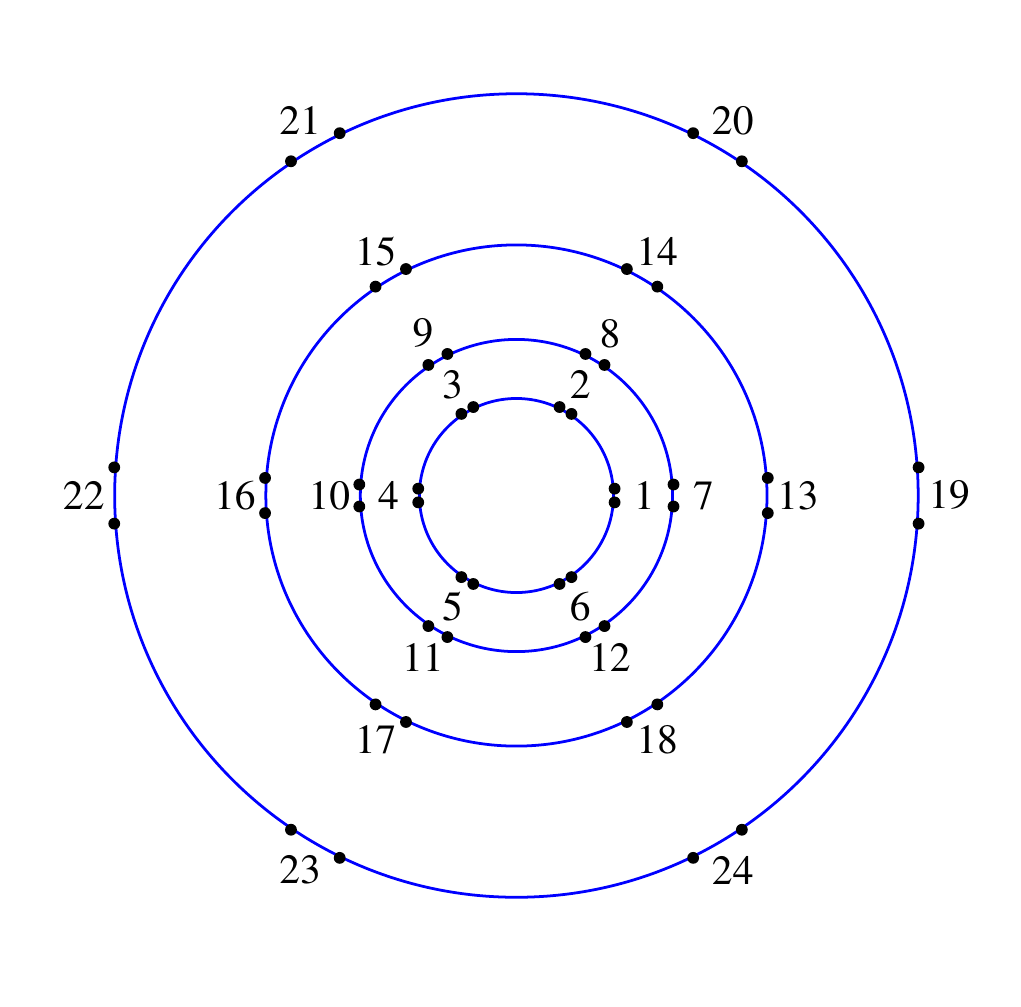}

    \caption{2D configuration corresponding to 48 $\sigma$ fields arranged on a cylinder with $N_x=6$ and $N_y=4$. We represent it on the plane according to the exponential map $z\mapsto \exp(i\theta)$. For clarity, we label the 24 physical spins, each one obtained from a pair of $\sigma$ fields. Using coordinates \eqref{phi2D}, spins with equal $y_n$ are located at the same radius.}
    \label{conf2D}
\end{figure}

Pairing function \eqref{gSC} is similar to the one obtained from $\ket{\psi_{ee}}$ using an OPE approximation \eqref{OPEpairingPlane}. This suggests that $\ket{\psi_{ee}}$ can be related to the weak pairing phase of \eqref{HpSC} as long as the OPE regime yields a good approximation of the CBs. The set of distances between the $\sigma$ fields can then be related to the chemical potential of the $p+ip$ superconductor. We expect then that $\ket{\psi_{ee}}$ can describe the topological weak pairing phase of \eqref{HpSC}, which has been associated to the Moore-Read Pfaffian state in the fractional quantum Hall effect \cite{ReadGreen, Miguel}.

Based on the previous analysis, we focus now 2D spin systems on finite cylinders. The lattice will contain $N_y$ spins along the longitudinal direction and $N_x$ spins along the periodic one. We follow the analysis presented in Sect. V for the OPE approximation on the cylinder. We set $z_n = \exp(i\theta_n)$ using the cylinder coordinates \eqref{cylcoord}, where $\phi_n$ corresponds to the location of the $n$-th physical spin and $\epsilon_n$ to the size of its reference pair. We parametrize them as (see Fig. \eqref{conf2D})
\begin{equation}
\phi_n = \frac{2\pi}{N_x} (x_n - i R y_n)
\label{phi2D}
\end{equation}
where $n=1,\cdots, N_x N_y$ labels the spin sites, $x_n \in \{1,\cdots N_x\}$ and $y_n \in \{1,\cdots, N_y\}$ are positive integers that define the lattice on the cylinder, and $R$ is the anisotropy factor. (We will use a regular square lattice, so we set $R=1$.) We also define the same separation for all reference pairs 
\begin{equation}
\epsilon_n = \frac{2\pi}{N_x}\epsilon.
\end{equation}
We can use complex values for $\epsilon$, but the radial ordering leads to subtleties when we extrapolate to the exact regime. We will focus then on real values, noting that the OPE regime corresponds to $0<\epsilon\ll 1$.

Using this notation, the OPE pairing function becomes
\begin{align}
g_{nm} = \frac{\sin\left(\frac{\pi}{N_x}\epsilon\right)} {2\sin\left(\frac{\pi}{N_x}(x_n-x_m - i R (y_n - y_m))\right)}.
\end{align}
Note that, for large values of $N_x$, we can approximate this expression by a power law, so the leading singular term is similar to \eqref{gSC}.

\begin{figure}[h]
  \centering

\includegraphics[width=1.0\linewidth]{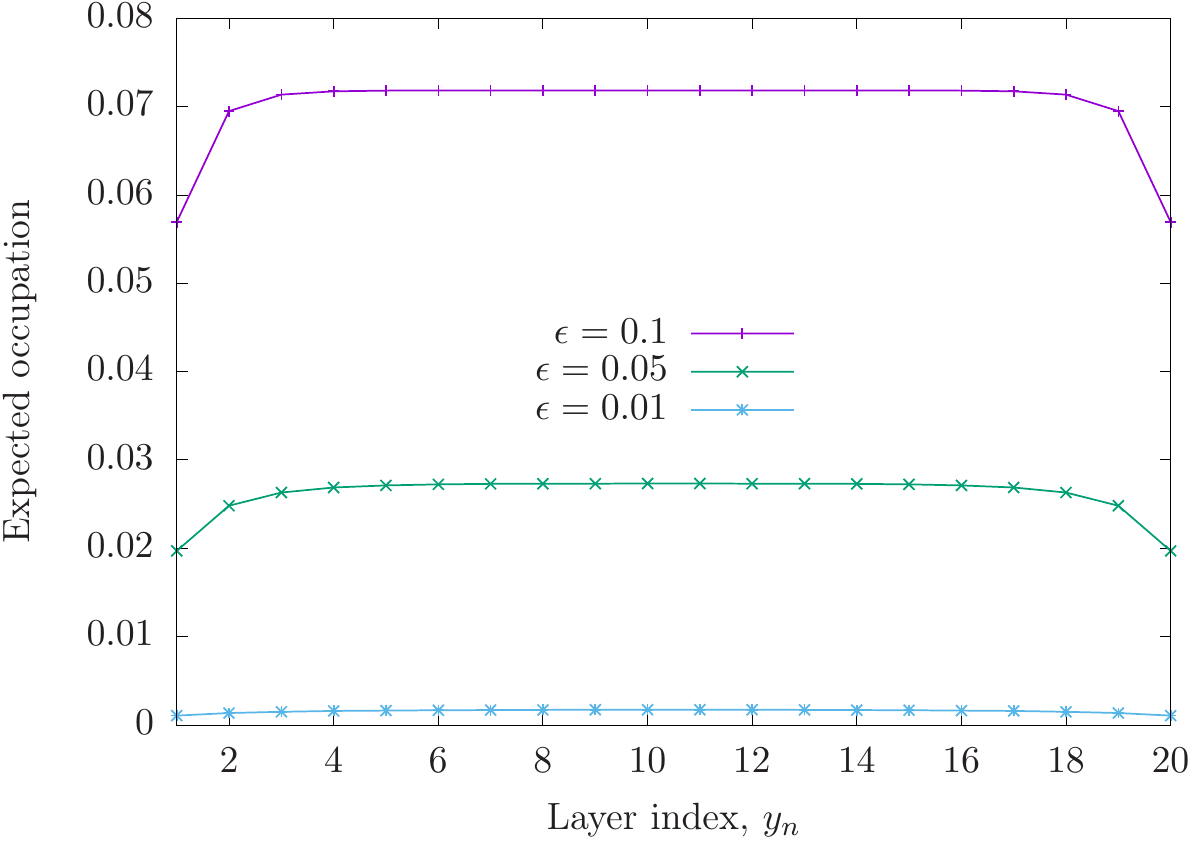}

    \caption{Expected occupation per site for different values of $\epsilon$ for a cylinder with $N_x=N_y=20$. The layer $y_n$ correspond to the longitudinal $y$-direction in the cylinder. Note that being periodic along the $x$-direction, the expectation value does not depend on $x_n$.}
    \label{dens2D}
\end{figure}

In order to characterize these wave functions, we study the entanglement spectrum \cite{LiHaldane} and the entanglement entropy obtained from the reduced density matrix $\rho_\text{cyl}$ of half a cylinder (corresponding to all sites with $y_n=1,\cdots,\frac{N_y}{2}$). Being a BCS state, we know that $\rho_\text{cyl}$ can be written as \cite{EntanglementSpinChains, Peschel}
\begin{equation}
\rho_\text{cyl} = \frac{1}{Z}\exp\left(\sum_m \lambda_m b^\dg_m b_m \right), 
\end{equation}
where $\{b_m\}$ are fermionic modes and $Z$ the normalization constant. In Appendix F, we describe a general algorithm to obtain both the spectrum $\{\lambda_m\}$ and the fermionic modes from the pairing function $g_{nm}$.

\begin{figure}[h]
\centering
\subfigure[]{
   \includegraphics[width=0.85\linewidth]{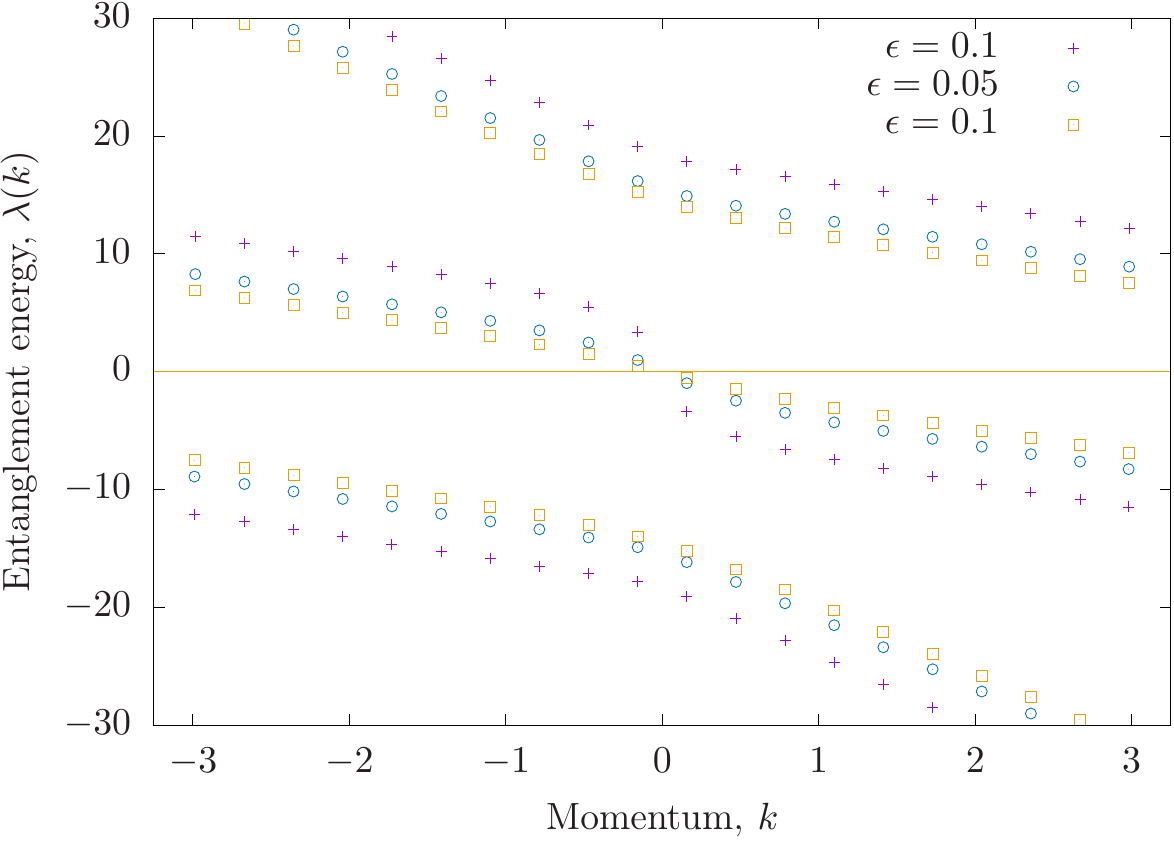}
   \label{entspec2D}
   }
\subfigure[]{
   \includegraphics[width=0.85\linewidth]{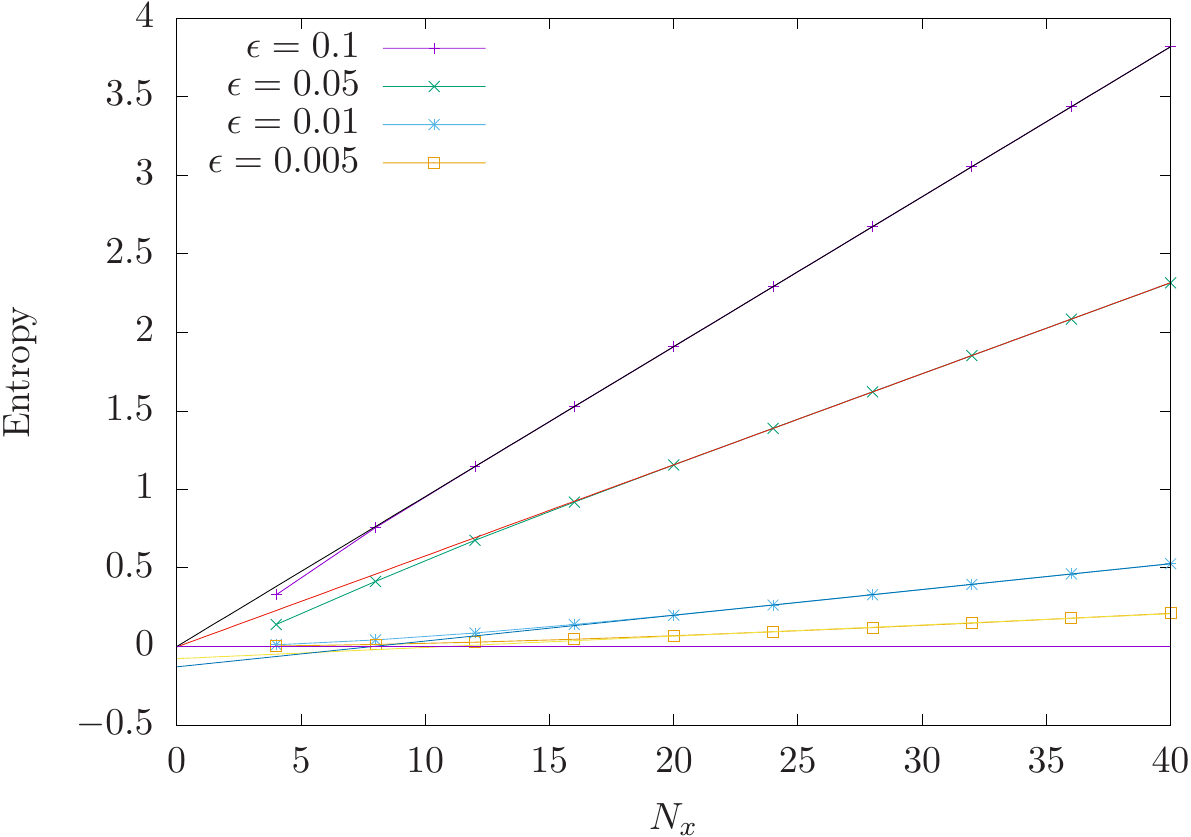}
   \label{ententr2D}
}
    \caption{(a) Entanglement (single-body) spectrum for different values of $\epsilon$ on a cylinder with $N_x=N_y=20$. (b) Scaling of the entanglement entropy as a function of the circumference of the cylinder for different values of $\epsilon$ and set $N_y=80$.}
\end{figure}

We compute the expected occupation per site for different values of $\epsilon$ (see Fig. \eqref{dens2D}). We see that the boundaries do not affect the physics deep inside the bulk for large enough $N_y$. Given that in the limit $\epsilon\to 0$ the state corresponds to a trivial vacuum, the occupation is small in the OPE regime.

The periodicity in the $x$-direction is preserved in $\rho_\text{cyl}$, so that we can associate a momentum $k$ to each mode. We can then write the single-body entanglement spectrum as a dispersion relation. In Fig. \eqref{entspec2D}, we see the single-body spectrum for different values of $\epsilon$. It corresponds to a chiral free fermion. For values close to $\epsilon\to 0$, there is a gap in the dispersion that closes at around $\epsilon\sim 0.1$. This behavior is in agreement with the entanglement spectrum of $p+ip$ superconductors in the weak pairing phase \cite{chiralfermion}.

From the entanglement spectrum, we computed the scaling of the entanglement entropy for different values of $\epsilon$ by changing the cincumference of the cylinder (see Fig. \eqref{ententr2D}). In all cases, the scaling follows an area law $S(N_x)\sim c N_x$, with non-universal slopes. According to the scaling, there is no topological correction in the entanglement entropy. Once again, this is in agreement with the behavior of $p+ip$ superconductors \cite{chiralfermion, triplettopSC}.


\section{Conclusions}

We have presented a characterization of many-body states for lattice systems constructed from the CBs of the chiral Ising CFT. The basic feature of the construction is the use of pairs of $\sigma$ fields to describe single localized spins. Writing these CBs using local vertex operators enables us to relate this formalism to usual matrix product states. This rewriting makes explicit the relation between the ancillary CFT degrees of freedom and the lattice fermionic modes.

We have provided evidence that states constructed from CBs using only $\sigma$ fields can be written as BCS states. A partial proof of this fact can be obtained whenever an OPE approximation is valid. In this case, an explicit BCS form can be obtained using the local vertex operator formalism. In the case of translationally invariant 1D configurations, we can go beyond the approximation and write a full non-perturbative proof. This also allows us to obtain a whole family of quasi-local parent Hamiltonians that can be written as quadratic fermionic forms. They are closely related to the critical ITF Hamiltonian \eqref{ITFh1}. In particular, we presented a proof that the ground state of the critical ITF Hamiltonian can be obtained exactly from this construction. The first excited state of the even-parity sector of this Hamiltonian can also be obtained using CBs with fermions in the asymptotic CFT states.

The OPE approximation can be used to study large 2D spin configurations. By placing the degrees of freedom on finite cylinders, we have related the states obtained from the CBs in the OPE regime to the weak pairing phase of the $p+ip$ superconductor. This has been done via the entanglement spectrum obtained from the reduced density matrix of half of the cylinder.

Further work is needed to deepen the connection between CBs and the ground states of finite systems. In the case of the Ising CFT, this would mean a general proof that $\ket{\psi_{ee}}$ describes a BCS wave function regardless of the coordinate configuration. A deeper understanding of the formalism may produce other physically relevant states, such as the ground state of the 1D odd parity sector of the ITF Hamiltonian, or vortices in 2D superconductors. In additional, generalizations to other rational CFTs, such as the Potts model or the $\mathbb{Z}_n$ model \cite{Tsvelik, QuantumGroupsCFT, diFrancesco}, are worth studying. Due to the algebraic constraints, we expect those constructions to be related to anyon chains \cite{goldenChain,AnyonTN} or parafermions \cite{para1,para2,para3}.


\section*{Acknowledgments}

We would like to thank H.-H. Tu for being part of this project at its early stages and for suggesting the algorithm for variational parent Hamiltonians. We would also like to thank A.E.B. Nielsen, J. Slingerland, and R. Sanchez for useful discussions. This work is supported by the Spanish Research Agency (Agencia Estatal de Investigaci\'on) through the grant IFT Centro de Excelencia Severo Ochoa SEV-2016-0597, and funded by Grant No. FIS2015-69167-C2-1-P from the Spanish government, and QUITEMAD+ S2013/ICE-2801 from the Madrid regional
government. SM is supported by the FPI-Severo Ochoa Ph.D. fellowship No. SVP-2013-067869.


\section*{Appendix A: Determinant form for the 1D wave function}

We can rewrite the wave function amplitudes as
\begin{equation}
\Psi_\pp = \frac{1}{\tilde N_0}\left(\sum_{\{s_1=1,\cdots\}}\tilde\epsilon_{\pp\sss}\,A(\{s_k\})\right)^{1/2}
\end{equation}
where
\begin{align}
A(\{s_k\}) =\bigg(& \prod_{j>i}^N \sin\left[\frac{\pi}{N}\left(j-i+\frac{1+2\delta}{4}(s_j-s_i)\right)\right]\\
&\sin\left[\frac{\pi}{N}\left(j-i-\frac{1+2\delta}{4}(s_j-s_i)\right)\right]\bigg)^{1/2}.\nonumber
\end{align}
Using the fact that (see Appendix in \cite{WFfromCB})
\begin{align}
\prod_{j>i}^N &\sin\left[\frac{\pi}{N}\left(j-i+\frac{\alpha}{4}(s_j-s_i)\right)\right] \\
&=\prod_{j>i}^N \sin\left[\frac{\pi}{N}\left(j-i-\frac{\alpha}{4}(s_j-s_i)\right)\right],
\end{align}
for arbitrary real $\alpha$, we can eliminate the square root in $A(\{s_k\})$ and lift the restriction on $s_1$ noting that
\begin{align}
\sum_{\{s_1=1,\cdots\}}\tilde\epsilon_{\pp\sss}\, A(\{s_k\})
=\frac{1}{2}\sum_{\{s_1=\pm1,\cdots\}}\tilde\epsilon_{\pp\sss}\, A(\{s_k\}).
\end{align}
Putting all the pieces together, we have
\begin{equation}
\Psi_\pp = \frac{1}{N_0}\left(\sum_{\{s_k\}}\tilde\epsilon_{\pp\sss}\,A(\{s_k\})\right)^{1/2},
\end{equation}
where now
\begin{equation}
A(\{s_k\}) = \prod_{j>i}^N \sin\left[\frac{\pi}{N}\left(j-i+\frac{1+2\delta}{4}(s_j-s_i)\right)\right]
\end{equation}
and
\begin{equation}
N_0^2 = 2 \tilde N_0^2 = 2^{N}\left(\frac{N}{2^{N-1}}\right)^{N/2}.
\end{equation}

Note first that, using the identities
\begin{equation}
\sum_{\sigma\in S_N}\text{sgn}(\sigma)\prod_{n=1}^N \alpha_n^{\sigma(n)-1} = \prod_{n>m}^N (\alpha_n-\alpha_m)
\end{equation}
and
\begin{equation}
\sin\left(\frac{\theta_n - \theta_m}{2}\right) = \exp\left(-i\frac{\theta_n + \theta_m}{2}\right)\left(\frac{z_n - z_m}{2i}\right),
\end{equation}
we have
\begin{align}
A(\{s_k\})&=\prod_{j>i}^N\left[ \exp\left(-i\frac{\theta_j + \theta_i}{2}\right)\left(\frac{z_j - z_i}{2i}\right)\right]\nonumber\\
&=C_N\sum_{\sigma\in S_N}\text{sgn}(\sigma)\left(\prod_{j=1}^N a_{j,\sigma(j)}\right)\left(\prod_{j=1}^N b_{j,\sigma(j)}\right)
\end{align}
where
\begin{equation}
(V)_{r,t}=a_{r,t} = \exp\left(i\frac{2\pi}{N}r(t-1)\right)
\label{Vandermonde}
\end{equation}
defines a Vandermonde matrix,
\begin{equation}
b_{r,t} = \exp\left(i\frac{\pi}{4N}s_r(1+2\delta)(2t-N-1)\right)
\end{equation}
contains all the dependence on $\{s_k\}$, and
\begin{equation}
C_N = (2i)^{-N(N-1)/2}e^{-i\frac{\pi}{2}(N^2-1)}
\end{equation}

Coming back to $\Psi_\pp$, we can now sum over the auxiliary spins $\{s_k\}$
\begin{align}
\sum_{\{s_k\}}&\tilde\epsilon_{\pp\sss}\,A(\{s_k\}) \\
&= C_N\sum_{\sigma\in S_N}\text{sgn}(\sigma)\prod_{j=1}^N a_{j,\sigma(j)} \left(\sum_{\{s_k\}}\tilde\epsilon_{\pp\sss}\prod_{j=1}^N b_{j,\sigma(j)}\right).\nonumber
\end{align}

We can perform the sum
\begin{align}
\sum_{\{s_k\}}\tilde\epsilon_{\pp\sss}\prod_{j=1}^N b_{j,\sigma(j)}=2^N\prod_{j=1}^N f_{j,\sigma(j)}.
\end{align}
where
\begin{equation}
f_{r,t} = \left\{
  \begin{array}{l l}
    \cos\left[\frac{\pi}{4N}(1+2\delta)\left(2t-N-1\right)\right], & \,\,m_r=0,\\
    i\sin\left[\frac{\pi}{4N}(1+2\delta)\left(2t-N-1\right)\right], & \,\, m_r=1,
  \end{array} \right.
\end{equation}
and we make use again of the pair-wise basis $\mm=(m1_,\cdots,m_N)$. In other words, we have a cosine whenever the $r$-th reference pair fuses to an identity and sine when it fuses to a fermion. We can now clean everything up. Note first that
\begin{align}
N_0^2 &=  2^{N}\prod_{j>i}^N \sin\left(\frac{\pi}{2N}\left(j-i\right)\right)\nonumber\\
&= 2^N C_N\sum_{\sigma\in S_N}\text{sgn}(\sigma)\left(\prod_{j=1}^N a_{j,\sigma(j)}\right) \\
&=  2^N C_N\det(V), \nonumber
\end{align}
where $V$ is the Vandermonde matrix defined by $a_{r,t}$. If we define the matrix $F_\mm$ by the elements $f_{r,t}$, we have
\begin{equation}
\Psi_\mm = \left(\frac{\det(F_\mm*V)}{\det(V)}\right)^{1/2},
\end{equation}
where $F_\mm*V$ is the Hadamard product
\begin{equation}
(F_\mm*V)_{r,t} = a_{r,t}f_{r,t}\,.
\end{equation}
%


\section*{Appendix B: BCS state from determinant form for 1D wave functions}

In order to show that the wave function amplitudes \eqref{psieedet} correspond to a BCS state, we need to write them as Pfaffians obtained from a given pairing function. We will accomplish this by using the multilineality of the determinant.

First, consider the matrix
\begin{equation}
(U)_{r,t} = \frac{1}{\sqrt N} \exp\left(i\frac{2\pi}{N}r\left(t-\frac{1}{2}\right)\right).
\end{equation}
It is easy to show that $U$ is a unitary matrix. Also, by multilineality of the determinant,
\begin{equation}
\Psi_\mm^2 = \frac{\det(F_\mm*V)}{\det(V)} = \frac{\det(F_\mm*U)}{\det(U)}.
\end{equation}
%
Now, note that we can write
\begin{equation}
\det(F_\mm*U) = C_N(\delta)^2 \det(H_\mm*U),
\end{equation}
where 
\begin{align}
C_N(\delta)^2 &= \prod_{m=1}^N \cos\left[\frac{\pi}{4N}(1+2\delta)\left(2m-N-1\right)\right],
\end{align}
and $H_\mm$ is defined by matrix elements
\begin{equation}
(H_\mm)_{r,t} = \left\{
  \begin{array}{l l}
    1, & \,\,m_r=0,\\
    i\tan\left[\frac{\pi}{4N}(1+2\delta)\left(2t-N-1\right)\right], & \,\, m_r=1.
  \end{array} \right.
\end{equation}

We can further simplify this expression. If we define $M_\mm = \left((H_\mm*U)U^\dg\right)$ and
\begin{align}
g_r &= (-1)^r\frac{i}{N}\sum_{m=1}^N  \tan\left[\frac{\pi(1+2\delta)}{4N}\left(2m-N-1\right)\right]e^{i\frac{2\pi}{N}mr}\nonumber\\
& =(-1)^r\frac{2}{N}\sum_{k>0}\tan\left(\frac{k}{4}\right)\sin\left(kr\right),
\label{BCSpairingfun}
\end{align}
(using momenta $k$ as defined in \eqref{kdef}), it is easy to see that
\begin{equation}
(M_\mm)_{r,t}  = \left\{
  \begin{array}{l l}
    \delta_{r,t}, & \,\,m_r=0,\\
    g_{r-t}, & \,\, m_r=1.
  \end{array} \right.
\end{equation}
Note that $g_r$ is an anti-symmetric function. Taking into account that $\sum m_n=2R$ is an even number, assume that the $1$'s are located at positions $r(1)<\cdots<r(2R)$. In order to compute the determinant of $M_\mm$, note that
\begin{equation}
\det(M_\mm) = \det(\textbf{G}_\mm)
\end{equation}
where
\begin{equation}
(\textbf{G}_\mm)_{ij} = g_{r(i)-r(j)},
\end{equation}
is the $2R\times 2R$ anti-symmetric matrix obtained from $M_\mm$ by keeping only the rows and columns corresponding to $r(1),\cdots,r(2R)$. Being anti-symmetric, note also that
\begin{equation}
\det(\textbf{G}_\mm) = \text{Pf}^2 (\textbf{G}_\mm).
\end{equation}

Summing up, we have
\begin{align}
\Psi_\mm^2 & = C_N(\delta)^2\frac{\det(H_\mm*U)}{\det(U)}\nonumber\\
& = C_N(\delta)^2\det\left(M_\mm\right)\\
& = \left(C_N(\delta)\,\text{Pf} (\textbf{G}_\mm)\right)^2.\nonumber
\end{align}
Given that this result holds for all $\mm$, we conclude that $\ket{\psi}=\sum_\mm \Psi_\mm\ket{\mm}$ corresponds to a BCS state defined by pairing function \eqref{BCSpairingfun}.


\section*{Appendix C: Finding the 1D parent Hamiltonians}

Consider a family of Hamiltonian terms
\begin{equation}
H_\alpha = \sum_{i_1,\cdots,i_k}h^{(\alpha)}_{i_1,\cdots,i_k}
\end{equation}
which can be either local or non-local. For convenience, we set $H_0 = \id$. Given a wavefunction $\ket{\Psi}$, we would like to find a linear superposition of these operators that will have $\ket{\Psi}$ as an eigenstate. In other words, we want to find coefficients  $J_\alpha$ such that
\begin{equation}
\left(\sum_\alpha J_\alpha H_\alpha\right) \ket{\Psi} = 0.
\label{superposition}
\end{equation}
In order to solve this, consider the matrix
\begin{equation}
(M)_{\alpha\beta} = \braket{\Psi\left|H_\alpha H_\beta\right|\Psi}.
\end{equation}
It is easy to see that condition \eqref{superposition} will be satisfied for a certain set of coefficients $\{J_\alpha\}$ if and only if $M$ has a non-trivial kernel. (Note that $M$ is positive-definite.)

Going back to the Hamiltonian terms \eqref{HamTerms1D}, we can use the JW transformation,
\begin{align}
\sigma_n^z &= 1-2c_n^\dg c_n, \nonumber\\ 
\sigma_n^x &= \prod_{m=1}^{n-1}(1-2c_m^\dg c_m)(c_m^\dg+c_m), \\
\sigma_n^y &= i \prod_{m=1}^{n-1}(1-2c_m^\dg c_m)(c_m^\dg-c_m),\nonumber
\end{align}
to obtain
\begin{align}
Z &\mapsto -\sum_{n}(1-2c_n^\dg c_n), \nonumber\\
X_{r} &\mapsto
- \sum_{n}(c_n^\dg - c_n)(c_{n+r}^\dg + c_{n+r}),\\
Y_{r} &\mapsto \sum_{n} (c_n^\dg + c_n)(c_{n+r}^\dg - c_{n+r}).\nonumber
\end{align}
We pick the convention for the JW transformation so that $\ket{0}_c = \ket{\uparrow}^{\otimes N}$. Note that, for the even parity sector $\braket{Q}=\braket{\prod_n \sigma_n^z}=1$, we have antiperiodic boundary conditions for the fermions $c_{N+m} = -c_{m}$.


It is illuminating to write these operators in terms of Majorana fermions
\begin{equation}
a_{2n-1} = c_n + c_n^\dg, \quad a_{2n} = \frac{c_n-c_n^\dg}{i}.
\end{equation}
Antiperiodic boundary conditions on the fermions imply $a_{2N+r} = -a_r$. In these variables, we have
\begin{align}
Z &= i\sum_n a_{2n-1}a_{2n},\nonumber\\
X_r &= i\sum_n a_{2n}a_{2(n+r)-1},\\
Y_r &= - i\sum_n a_{2n-1}a_{2(n+r)}.\nonumber
\end{align}
%

Now, let us consider the variational Hamiltonian family defined in \eqref{HamFamily1D}. Using the Majorana variables, we obtain
\begin{align}
\tilde H_1 &= i\sum_n a_n a_{n+1},\\
\tilde H_r &= i \sum_n (-1)^n a_n a_{n+2(r-1)-1},\nonumber
\end{align}
with $r = 2,\cdots, N/2+1$. Tha action of the KW transformation on the Majorana fermions is simply
\begin{equation}
a_r \mapsto a_{r+1},
\end{equation}
which acts as expected on the Hamiltonian family. Note that this action mimics the interpretation of the KW transformation as a consequence of the braiding of sigma fields as discussed in Ref.\cite{WFfromCB}.

The Hamiltonian family we have used is very similar to the conserved quantities of the Ising model, seen as an integrable model \cite{SKoriginal}. Those can be obtained from
\begin{equation}
E_p = (-1)^p\frac{i}{2p}\sum_n a_n a_{n+p}.
\end{equation}
Note that $\tilde H_1 = -2E_1$ and that $[E_p,E_q]=0$. It has been shown that formal manipulations of $\{E_p\}$ can yield lattice representations of the Virasoro algebra \cite{SKoriginal, SKnonintegrable}.


\section*{Appendix D: Towards a generalized Wick theorem}

The OPE of two $\sigma$ fields is given by \eqref{fullSigmaOPE}. One can easily
identify the fields $\alpha(z)$ as the ones appearing  in the fusion channel of the CVOs $V_{00}$
and $V_{\chi \chi}$, while the fields $\beta(z)$ are the ones  appearing in the CVOs $V_{0 \chi}$ and $V_{\chi0}$. 
This implies that \eqref{N4VertexCondition} holds provided the fields $\alpha(z_i)$
and $\beta(z_i)$ satisfy the relation 
\begin{align}
\braket{ \alpha_1 \alpha_2 \alpha_3 \alpha_4 } \braket{ \beta_1 \beta_2 \beta_3 \beta_4  }  = & \braket{  \beta_1 \beta_2   \alpha_3 \alpha_4  } \braket{ \alpha_1 \alpha_2   \beta_3 \beta_4 } \label{generalizedWick} \\
&- \braket{ \beta_1 \alpha_2 \beta_3 \alpha_4    } \braket{ \alpha_1 \beta_2 \alpha_3 \beta_4    }\nonumber \\
&+  \braket{ \beta_1 \alpha_2 \alpha_ 3 \beta_4   } \braket{ \alpha_1 \beta_2 \beta_3 \alpha_4 },\nonumber
\end{align} 
where $\alpha_i = \alpha(z_i)$ and $\beta_i = \beta(z_i)$. This equation coincides with the standard Wick theorem if $\alpha(z)=\id$ and $\beta(z) = \chi(z)$. Let us provide other examples.

Suppose $\alpha_1 = T(z_1)$ with $T(z)$ the stress tensor \cite{diFrancesco}, $\alpha_2 = \alpha_3 = \alpha_4 = \id$ and $\beta_i = \chi(z_i)$. Using this, \eqref{generalizedWick} becomes
\begin{align}
\braket{ T_1 } \braket{ \chi_1 \chi_2 \chi_3 \chi_4  }  = & 
\braket{  \chi_1 \chi_2   } \braket{ T_1   \chi_3 \chi_4 } - \braket{ \chi_1  \chi_3    } \braket{ T_1 \chi_2 \chi_4  } \nonumber\\
&+  \braket{ \chi_1  \chi_4 } \braket{ T_1 \chi_2 \chi_3  }.
\end{align}
The left hand side of the equation vanishes because on the plane $\braket{ T(z) } = 0$. To find $\braket{ T \chi \chi }$, we use Ward identities \cite{BPZ} to conclude
\begin{equation}
\braket{ T(z_1) \chi({z_2}) \chi(z_3) } = \frac{ z_{23}}{ 2 z_{12}^2 z_{13}^2}.
\end{equation}
Plugging these equations into \eqref{generalizedWick} yields
\begin{align}
\frac{1}{ z_{12}} \frac{ z_{34}}{ 2 z_{13}^2 z_{14}^2}  &-  \frac{1}{ z_{13}} \frac{ z_{24}}{ 2 z_{12}^2 z_{14}^2} 
+ \frac{1}{ z_{14}} \frac{ z_{23}}{ 2 z_{12}^2 z_{13}^2} = \\
&\frac{ z_{12}  z_{34} - z_{13} z_{24} + z_{14} z_{23}}{ 2 (z_{12} z_{13} z_{14})^2} =0, \nonumber
\end{align}
so that the condition is satisfied.

As a more elaborate example, choose $\alpha_i = \id$, $\beta_1 = L_{-n} \chi(z_1)$ with $L_{-n}$ the mode operator of the stress tensor that belongs to the representation of the Virasoro algebra \cite{diFrancesco}, and $\beta_i(z) = \chi(z_i) \; (i=2,3,4)$. Equation \eqref{generalizedWick} becomes
\begin{align}
\braket{ \left(L_{-n}\chi_1\right) \chi_2 \chi_3 \chi_4  }  = & 
\braket{  \left(L_{-n}\chi_1\right) \chi_2   } \braket{  \chi_3 \chi_4 } \\
&- \braket{\left(L_{-n}\chi_1\right)  \chi_3    } \braket{ \chi_2 \chi_4  } \nonumber\\
&+  \braket{ \left(L_{-n}\chi_1\right)  \chi_4 } \braket{ \chi_2 \chi_3  },\nonumber
\end{align}
where 
\begin{equation}
 L_{-n} \chi_1(z_1) = \oint_{z_1} d \zeta  \; (\zeta - z_1)^{ - n +1} \; T(\zeta) \, \chi(z_1) , \quad n \geq 1
\end{equation}
(We have suppressed the denominator $2 \pi i$ in the integral.) Equation \eqref{generalizedWick} can be written as
\begin{equation}
\Omega_n \equiv \oint_{z_1} d \zeta  \; (\zeta - z_1)^{ - n +1} \; f(\zeta, \{ z_i \} ) = 0,
\label{OmegaCondition}
\end{equation}
with
\begin{align}
 f(\zeta, \{ z_i \} ) =& \braket{T(\xi)\chi_1\chi_2\chi_3\chi_4}- \braket{T(\xi)\chi_1\chi_2}\braket{\chi_3\chi_4}\\
 &+ \braket{T(\xi)\chi_1\chi_3}\braket{\chi_2\chi_4} - \braket{T(\xi)\chi_1\chi_4}\braket{\chi_2\chi_3}.\nonumber
\end{align}
We now use the familiar identity for general fields $\phi_i$ with conformal weights $h_i$ \cite{diFrancesco}
\begin{align}
&\braket{T(\zeta) \prod_{i} \phi_i(z_i) } =\\
&\qquad \left[   \sum_{i}  \left(   \frac{ h_i}{(\zeta- z_i)^2 } + \frac{1}{ \zeta - z_i}  \frac{ \partial}{ \partial z_i}  \right)  \right] 
\braket{ \prod_{i} \phi_i({z_i})}, \nonumber
\end{align}
to find
\begin{equation}
 f(\zeta, \{ z_i \} ) =  \left( \frac{ \zeta - z_1}{  (\zeta - z_2) (\zeta- z_3) (\zeta - z_4)} \right)^2   \frac{ z_{23} z_{24} z_{34}}{ 2 z_{12} z_{13} z_{14}}.
\end{equation}
Hence, equation \eqref{OmegaCondition} becomes
\begin{align}
\Omega_n &=  \oint_{z_1} d \zeta  \; (\zeta - z_1)^{ - n +1} \;  \left( \frac{ \zeta - z_1}{  (\zeta - z_2) (\zeta- z_3) (\zeta - z_4)} \right)^2 \nonumber\\ 
&= \oint_{z_1} d \zeta  \;   \frac{ (\zeta - z_1)^{ -n + 3}}{  [ (\zeta - z_2) (\zeta- z_3) (\zeta - z_4)]^2}  = 0. \quad (n\geq 1)
\end{align}
This equation holds for $n=1,2,3$ but for $n=4$ one has
\begin{equation}
\Omega_4 = \oint_{z_1} d \zeta  \;   \frac{ (\zeta - z_1)^{ -1}}{  [ (\zeta - z_2) (\zeta- z_3) (\zeta - z_4)]^2}  =  \frac{ 1}{ (z_{12} z_{13} z_{14})^2} 
\end{equation}
It seems that $\Omega_n \neq 0$ for $n \geq 4$. Hence in these cases \eqref{generalizedWick} does not hold. What is the explanation of this fact? 

The characters of the Verma modules $\V_\id$ and $\V_\chi$ are given by
\begin{align}
\chi_0(q)  &=  {\rm Tr}_{\V_\id}   q^{ L_0 - c/24} \\
&=q^{ - \frac{1}{48}} \left(  1 + q^2 + q^3 + 2 q^4 + 2 q^5 + 3 q^6 + \dots   \right), \nonumber\\
\chi_1(q) & =  {\rm Tr}_{\V_\chi}   q^{ L_0 - c/24}\\
&= q^{ - \frac{1}{48}- \frac{1}{2} } \left(1 + q + q^2 + q^3 + 2 q^4 + 2 q^5 + \dots \right).  \nonumber 
\end{align}
Notice that at level $n=4$ there are two states in the Majorana sector. As a matter of fact, the descendants we had considered above correspond to the derivatives of the field $\chi(z)$,
\begin{equation}
(L_{-n}  \chi)(0) = \frac{n+1}{2} \chi_{- n - \frac{1}{2}} 
\end{equation}
The conclusion is that equation \eqref{N4VertexCondition} reduces to equation \eqref{generalizedWick} only if the fields $\alpha$ and $\beta$ that appear in the OPE \eqref{fullSigmaOPE} are unique at a given level. Otherwise one has to consider all the fields appearing at the same level.


\section*{Appendix E: Fourier transform of 2D pairing function}

If $\mu>0$, we have $\xi_\kk>0$ for small momenta and
\begin{equation}
g_\kk \sim -\frac{2\mu}{\hat\Delta(k_x+ik_y)}.
\end{equation}
Let us try to fix the constants in the Fourier transform, at least in an asymptotic way. Taking $L=aN$ to be the length of the systems (so that the total number of sites is $N\times N$), we can define
\begin{equation}
k_x = \frac{2\pi(n-\frac{1}{2})}{aN}, \quad k_y = \frac{2\pi(m-\frac{1}{2})}{aN},
\end{equation}
where $n,m=-N/2,-N/2+1\cdots,N/2$. Using this, we have
\begin{align}
\frac{1}{N^2}&\sum_{k_x,k_y} \frac{\exp\left[i\left(k_x x+k_y y\right)\right]}{k_x+ik_y} \\
&\to \frac{a^2}{(2\pi)^2}\int_{-\pi/a}^{\pi/a}dk_x\int_{-\pi/a}^{\pi/a}dk_y\frac{\exp\left[i\left(k_x x+k_y y\right)\right]}{k_x+ik_y}.\nonumber
\end{align}
Here we need to be careful. We will both take the limit $a\to 0$ and keep it explicitly in the prefactor. (This can be fixed by changing the normalization of the Fourier transform.) Note that for $y>0$
\begin{equation}
\int_{-\infty}^\infty \frac{dk_y}{2\pi i} \frac{\exp\left[i\left(k_x x+k_y y\right)\right]}{k_y-ik_x} = \Theta(k_x) \exp\left[ik_x\left(x+i y\right)\right].
\end{equation}
Using this, we have
\begin{equation}
g(\rr)\to -\frac{2a^2 \mu}{2\pi i \hat\Delta}\frac{1}{x+iy}.
\end{equation}
For a fixed number of fermions, this corresponds to the Moore-Read state for the FQHE. In this phase, the ground-state of the $p+ip$ conductor is then a grand-canonical state of fermions with this pairing.


\section*{Appendix F: Bogoliubov transformation from a BCS pairing matrix}

Let us consider a fermionic system with on-site creation operators $c^\dagger_i$, $i\in \{1,\cdots, N\}$ and annihilation operators $c_i$. We will adopt the following notation:
\begin{equation}
\{C_l\}_{l=1}^{2N}=\{c_1,\cdots,c_N,c^\dagger_1,\cdots,c^\dagger_N\}.
\end{equation}
Thus, creation and annihilation operators are bundled together. Let us consider a different set of creation and annihilation operators,
\begin{equation}
\{B_l\}_{l=1}^{2N}=\{b_1,\cdots,b_N,b^\dagger_1,\cdots,b^\dagger_N\}
\end{equation}
with $B_l=\sum_p M_{lp} C_p$. The linear transformation will be a Bogoliubov transformation if the $b^\dagger$ and $b$ are bona-fide creation and annihilation operators, with the expected anticommutation and adjoint relations. The first condition is that $M$ is unitary. If that is the case, the Bogolibov matrix $M$ can be naturally split:
\begin{equation}
\begin{pmatrix}b \\ b^\dagger\end{pmatrix} =
\begin{pmatrix}D & E \\ E^* & D^*\end{pmatrix}
\begin{pmatrix}c \\ c^\dagger\end{pmatrix},
\end{equation}
where $D$ and $E$ are $N\times N$ complex matrices, $D^*$ and $E^*$are their complex conjugates (not Hermitian adjoints!) and they must fulfill
\begin{align}
DD^\dagger + EE^\dagger = \id, \quad DE^T + ED^T = 0
\end{align}
so that matrix $M$ will be unitary. Notice that $A^\dagger$ is the Hermitian adjoint, and $A^T$ is merely the transpose.

In our case, the BCS state is defined via the pairing function
$g_{ij}$, which is anti-symmetric, $g_{ij}=-g_{ji}$,
\begin{equation}
\ket{\Psi}=\exp\left( \sum_{ij} g_{ij} c^\dg_i c^\dg_j \right) \ket{0}_c \equiv \exp(P) \ket{0}_c
\end{equation}
where the last relation defines the pairing operator $P$. This state is the vacuum of a certain Bogoliubov set of operators, $\{B_l\}_{l=1}^{2N}=\{b^\dagger_1,\cdots,b^\dagger_N,b_1\cdots,b_N\}$, which means that
\begin{equation}
b_k\ket{\Psi}=0, \qquad k\in \{1,\cdots,N\}.
\end{equation}

Let us impose that condition in order to find the Bogoliubov transformation $M$. By definition, 
\begin{equation}
b_k = \sum_i D_{ki} c_i + E_{ki}c^\dagger_i,
\end{equation}
so our condition becomes
\begin{equation}
0 = b_k \exp(P) \ket{0} = \left( \exp(P) b_k +[b_k,\exp(P)] \right)\ket{0}.
\end{equation}
Remember that $b_k$ can be expanded as a linear combination of $c_i$ and $c^\dagger_i$. Using
\begin{equation}
[c_i,f(\{c_j,c^\dagger_j\})]=\frac{\partial f}{\partial c^\dagger_i},
\end{equation}
we find that
\begin{equation}
[c_i,\exp(P)]=\exp(P) \left(\sum_j g_{ij}c^\dagger_j \right).
\end{equation}
Of course, $c^\dagger_i$ commutes with $\exp(P)$. Thus, the annihilation condition becomes
\begin{equation}
\exp(P) \left\{ \sum_i D_{ki}c_i +  \sum_{ij}D_{ki}g_{ij}c^\dagger_j+ \sum_i E_{ki}c^\dagger_i \right\} \ket{0} =0.
\end{equation}
which implies the following relation between $D$, $E$ and $g$:
\begin{equation}
\sum_i D_{ki} g_{ij} + E_{kj}=0.
\end{equation}
Thus, in order to find the Bogoliubov transformation given the pairing matrix $g$, we have to solve the following matrix equations:
\begin{align}
Dg+E &=0, \nonumber\\
D D^\dg + E E^\dg &=\id, \\
DE^T + ED^T&= 0,\nonumber
\end{align}

From the first equation we get $E=-Dg$, which when inserted into the third equation yields $D(g^T+g)D=0$. But this relation is trivial due to the antisymmetry of $g$.
Then, the only non-trivial equation becomes
\begin{equation}
D \left( \id + g g^\dagger \right) D^\dagger = \id.
\end{equation}
This equation can be easily solved in the eigenbasis of $\id + g g^\dg$, which is self-adjoint and positive-definite.



\begin{thebibliography}{10}
\providecommand{\url}[1]{\texttt{#1}}
\providecommand{\urlprefix}{URL }
\providecommand{\eprint}[2][]{\url{#2}}

\bibitem{Onnes}
H. K. Onnes, Commun. Phys. Lab. Univ. Leiden \textbf{12}, 120 (1911).

\bibitem{BCSoriginal}
J. Bardeen, L. N. Cooper, J. R. Schrieffer, Phys. Rev. \textbf{108}, 1175 (1957).

\bibitem{Onsager}
L. Onsager, Phys. Rev. \textbf{65}, 117 (1944).

\bibitem{Henkel}
M. Henkel, \textit{Conformal invariance and critical phenomena}, Springer, 1999.

\bibitem{Sachdev}
S. Sachdev, \textit{Quantum Phase Transitions}, Cambridge University Press, Cambridge, 1999.

\bibitem{Mussardo}
G.  Mussardo,
\textit{Statistical  Field  Theory}, Oxford University Press, 2009.

\bibitem{ReadGreen}
N. Read, D. Green,
Phys. Rev. B \textbf{61}, 10267 (2000).

\bibitem{Miguel}
M. Iba\~nez, J. Links, G. Sierra, S.-Y. Zhao,
Phys. Rev. B \textbf{79}, 180501(R) (2009).
C. Dunning, M. Iba\~nez, J. Links, G. Sierra, S.-Y. Zhao,
J. Stat. Mech. P08025 (2010).

\bibitem{MooreRead91}
G. Moore, N. Read,
Nucl. Phys. B \textbf{360}, 362 (1991).

\bibitem{NayakAnyons}
 C. Nayak et al.,
 Rev. Mod. Phys. \textbf{80}, 1083 (2008).
 
 
\bibitem{BPZ}
A. Belavin, A. Polyakov, A. Zamolodchikov,
Nucl.Phys. \textbf{B241}, 33 (1984).

\bibitem{Tsvelik}
A.M. Tsvelik,
\textit{Quantum Field Theory in Condensed Matter Physics}, Cambridge University Press, 1995.

\bibitem{QuantumGroupsCFT}
C. G\'omez, M. Ruiz-Altaba, G. Sierra,
\textit{Quantum groups in two-dimensional physics}, Cambridge University Press, 1996.

\bibitem{diFrancesco}
P. di Francesco, P. Mathieu, D. S\'en\'echal,
\textit{Conformal field theory}. Springer, 1997.

\bibitem{Gogolin}
A. O. Gogolin, A. A. Nersesyan, A. M. Tsvelik,
\textit{Bosonization Approach to Strongly Correlated Systems}, Cambridge University Press, 1999.

\bibitem{iMPS}
J. I. Cirac, G. Sierra,
Phys. Rev. B \textbf{81}, 104431 (2010).

\bibitem{iMPS2}
A. E. B. Nielsen, J. I. Cirac, G. Sierra,
J. Stat. Mech. P11014 (2011). H.-H. Tu, A. E. B. Nielsen, J. I. Cirac, G. Sierra, New J. Phys. \textbf{16}, 033025 (2014). I.  Glasser, J. I.  Cirac, G. Sierra, A. E. B. Nielsen, Nucl. Phys. \textbf{B886}, 63 (2014). H.-H. Tu, A. E. B. Nielsen, G.  Sierra, Nucl. Phys. \textbf{B886}, 328 (2014). R. Bondesan, T. Quella,  Nucl. Phys. \textbf{B886}, 483 (2014). B.  Herwerth, G.  Sierra, Hong-Hao Tu, A.  E. B. Nielsen, Phys. Rev. B \textbf{91}, 235121 (2015). H.-H. Tu, G. Sierra, Phys. Rev. B \textbf{92}, 041119(R) (2015). I.  Glasser, J. I.  Cirac, G. Sierra, A. E. B. Nielsen, New J. Phys. 17, 082001 (2015).

\bibitem{KL}
A. E. B. Nielsen, J. I. Cirac, G. Sierra,
Phys. Rev. Lett. \textbf{108}, 257206 (2012). A. E. B. Nielsen, G. Sierra, J. I.  Cirac, Nature Communications \textbf{4}, 2864 (2013). B.  Herwerth, G.  Sierra, H.-H.  Tu, J. I. Cirac, A.  E. B. Nielsen, Phys. Rev. B \textbf{92}, 245111 (2015). I. Glasser, J.I. Cirac, G. Sierra, A. E. B. Nielsen, arXiv:1609.02435 (2016).

\bibitem{FQH-MPS}
B. Estienne, Z. Papi\'c, N. Regnault, B. A. Bernevig,
Phys. Rev. B \textbf{87}, 161112 (2013).

\bibitem{WFfromCB}
S. Montes, J. Rodr\' iguez-Laguna, H.-H. Tu, G. Sierra, Phys. Rev. B \textbf{95}, 085146 (2017).

\bibitem{LiHaldane}
H. Li, F.D.M. Haldane, Phys. Rev. Lett. \textbf{101}, 010504 (2008).

\bibitem{MooreSeiberg}
G. Moore, N. Seiberg,
Comm. Math. Phys. \textbf{123}, 177 (1989).

\bibitem{MooreSeibergNotes}
G. Moore, N. Seiberg,
"Lectures on RCFT." in \textit{Physics, Geometry and Topology}, Springer, 1990.

\bibitem{NayakWilczek}
C. Nayak, F. Wilczek,
Nucl. Phys. B \textbf{479}, 529 (1996).

\bibitem{IsingCB}
E. Ardonne, G. Sierra, 
J. Phys. A \textbf{43}, 505402, 2010.

\bibitem{topDefIsing}
D. Aasen, R.S. Mong, P. Fendley,
J. Phys. A: Math. Theor. \textbf{49}, 354001 (2016).

\bibitem{CardyContent}
J.L. Cardy,
Nucl. Phys. \textbf{B270}, 186 (1986).

\bibitem{ConformallyInvariantLimit}
Y.A. Bashilov, S.V. Pokrovsky,
Commun. Math. Phys. \textbf{113}, 115 (1987).

\bibitem{SKoriginal}
W. M. Koo, H. Saleur,
Nucl. Phys. \textbf{B426}, 459 (1994).

\bibitem{SKnonintegrable}
A. Milsted, G. Vidal, arXiv:1706.01436.

\bibitem{EntanglementSpinChains}
J. I. Latorre, E. Rico, G. Vidal,
Quant. Inf. Comp. \textbf{4}, 48 (2004).

\bibitem{Peschel}
I. Peschel, J. Stat. Mech. P06004 (2004).

\bibitem{chiralfermion}
N.  Bray-Ali,  L.  Ding,  S.  Haas,
Phys.  Rev.  B \textbf{80},180504(R) (2009).

\bibitem{triplettopSC}
T. P. Oliveira, P. Ribeiro, P. D. Sacramento,
Journal of Physics: Condensed Matter \textbf{26}, 425702 (2014).

\bibitem{goldenChain}
 A.   Feiguin et al.,
 Phys.  Rev.  Let. \textbf{98}, 160409 (2007).

\bibitem{AnyonTN}
R.N.C. Pfeifer, P. Corboz, O. Buerschaper, M. Aguado, M. Troyer, G. Vidal,
Phys. Rev. B \textbf{82}, 115126 (2010).

\bibitem{para1}
E. Fradkin and L. P. Kadanoff,
Nucl. Phys. B \textbf{170}, 1 (1980).

\bibitem{para2}
A.B. Zamolodchikov and V.A. Fateev,
Sov. Phys. JETP \textbf{62}, 215 (1985).

\bibitem{para3}
P. Fendley, J. Stat. Mech. p. P11020 (2012).

%
%
%
%
%
%

%


\end{thebibliography}
\end{document}